# Zoeppritz equations: from seismology to medical exploration

Harry G. Saavedra[1] and Ramiro Moro[1,2]

[1] Centro de Investigación en Bioingeniería, Universidad de Ingeniería y Tecnología (UTEC) 15063 Lima, Perú
[2] Department of Chemistry Physics and Engineering, Cameron University, Lawton, Oklahoma 73505, USA

**Resumen**

Hace más de un siglo, Karl Bernhard Zoeppritz derivó las ecuaciones que determinan los coeficientes de reflexión y transmisión en una interfaz plana para una onda sísmica incidente. Los coeficientes obtenidos son una función de los parámetros elásticos de los medios a cada lado de la interfaz y del ángulo de incidencia. Se han propuesto y utilizado aproximaciones de las ecuaciones en la exploración geofísica; sin embargo, el uso completo de las ecuaciones y su generalización a múltiples capas podrían ofrecer información más rica sobre las propiedades de los medios y ser útiles en el diagnóstico médico a través de ultrasonido. En este trabajo, investigamos cómo extraer información de los coeficientes de reflexión dependientes del ángulo, incluidos los ángulos críticos y la distorsión de la onda en la interfaz entre dos y tres medios. Se demuestra que es posible separar el efector de la densidad de la discrepancia en la velocidad del sonido al medir las amplitudes como función del ángulo de incidencia (AVA). Y el examen del ángulo crítico y la distorsión de la onda reflejada puede revelar el espesor de una capa intermedia, incluso con resolución menor a la longitud de onda. Estos estudios se podrían integra en imágenes médicas y servir para entrenar a sistemas de inteligencia artificial para asistir en diagnósticos. En particular, ellos podrían prevenir accidentes cerebro-vasculares por medio de la detección temprana de la formación y endurecimiento de placa en las arterias que irrigan el cerebro.

***Descriptores:*** *Ecuaciones de Zoeppritz, sismología, imágenes biomédicas, ultrasonido, diagnóstico médico*

**Abstract**

More than a century ago, Karl Bernhard Zoeppritz derived the equations that determine the reflected and transmitted coefficients at a planar interface for an incident seismic wave. The coefficients so obtained are a function of the elastic parameters of the media on each side of the interface and the angle of incidence. Approximations of the equations have been proposed and used in geophysical exploration, however, full use of the equations and their generalization to multiple layers could offer richer information about the properties of the media and be helpful in medical diagnosis via ultrasound. In this work, we investigate how to extract information from the angle-dependent reflection coefficients, including critical angles and the wave distortion at the interface between two and three media. It is shown that it is possible to separate the effect of density from speed of sound mismatch by measuring amplitudes as a function of angle of incidence (AVA). And examining the critical angle and waveform distortion of the reflected waves can reveal the thickness of an intermediate layer, even with subwavelength resolution. These studies could be integrated into medical imaging and also into the training of artificial intelligence systems that assist in diagnosis. In particular, they could help prevent cerebrovascular accidents by early detection of the formation and hardening of plaque in the arteries that irrigate the brain.

## 1. Introduction

In 1919, K. B. Zoeppritz's paper on the reflection of seismic waves was published [1-3, Appendix 1, Appendix 2]. In that work, it was established how to determine the amplitudes of reflected and transmitted waves at an interface between two homogeneous media by solving a set of four equations and Snell's law. This in principle allows to resolve the inverse problem of calculating the elastic constants by examining the amplitudes of the reflected waves as a function of incident angle. Since the Zoeppritz equations (Z.E.s) were considered complex, approximations were developed to gain insight for applications in geophysical exploration [4], where the amplitude versus angle (or offset) is studied [5-7]. With present technology, there should not be an obstacle to use the equations without approximations and infer information from the amplitudes, critical angles, and distortion of the wave at the interface. The data taken at different angles could improve medical exploration and diagnosis using ultrasound imagining. For example, prevention of cerebrovascular accidents (CVAs) produced by dislodging of plaque in the carotid arteries. These pieces of plaque travel in the blood flow and either obstruct an artery or produce a hemorrhage [8-9]. Interestingly, research has shown that, in addition to size, plaque composition is also an important risk factor [8]. Plaque composition in cardiac arteries has been determined employing spectral analysis of invasive ultrasound scattering [10]; however non-invasive studies of the carotid have the additional challenge of attenuation due to the intervening tissue. Another potential application of Z.E.s is in cancer diagnostics by detecting variations of sound wave velocities [11].

## 2. Methodology

The procedure starts by defining the fields of the incident wave and the reflected and transmitted ones. Then, the constraints due to continuity of displacements and stresses are written down and simplified with the help of Snell's law. Different however with respect to light, acoustic waves can have three polarizations when traversing in a solid: P waves are longitudinal pressure waves, the ones familiar for sound in fluids, S waves are transversal shear waves in two polarizations with displacements perpendicular to the plane of interface ($S_h$ waves) or parallel to it ($S_v$ waves). Also different from light, P and $S_v$ waves generate both P and $S_v$ transmitted and reflected waves when their angle of incidence is greater than zero. Appendix 1 is a translation of the original paper by Zoeppritz [1], which derives the equations for the partition of a wave at an interface between two media. Appendix 2 shows the derivation of the Z.E.s using modern nomenclature for longitudinal and transversal incident waves. In appendix 3, the equations for two interfaces between three different media are derived. This is an important practical problem that was not solved in the original paper.

## 3. Results

### 3.1 Amplitude vs. Angle (AVA)

To get insight in what the equations can reveal about the media that they are traversing, we can examine the Amplitude vs Angle (AVA). With reference to figures 1a and A2.1, we consider an incident P-wave impinging on an interface at an angle of incidence zero, in which case the reflection coefficient (ratio between amplitudes) is given by

$$\frac{A_1}{A_0} = \frac{z_2 - z_1}{z_2 + z_1} \qquad \text{Equation 1}$$

Where $A_0$ and $A_1$ are the incident and reflected amplitudes and $z_1$ and $z_2$ are the acoustic impedances defined as the product of density ($\rho$) and speed of sound ($\alpha$) for pressure waves (P-waves), $z = \rho\alpha$. Figure 1a shows the reflection coefficient ($A_1/A_0$) for typical values found in medical ultrasound and can be generated with the octave script attached. If this is the only information we obtain from the reflected waves, it is not possible to separate the effect of density mismatch from velocity mismatch between the two media. In addition, when evaluating amplitudes, the intensity attenuation is difficult to estimate due to the differences in travel distance, therefore, deriving acoustic impedances based solely on the amplitude of the reflected waves is not reliable. Nevertheless, the simple result in eq. 1 can be a diagnostic tool if other parameters are known, for example in the study of plaque accumulated in the heart arteries by invasive ultrasound [10, 12]. It is possible to take advantage of the dependence of the amplitude vs angle of incidence (AVA) to infer additional information. We refer to figure 1b where the acoustic impedance is identical for two media, which means that there is no reflection at an angle of incidence zero. However, a larger $\alpha$ in the second medium produces a reflected P-wave with positive polarity at larger angles of incidence (blue lines) while a slower $\alpha$ produces inverted polarity (red lines). This demonstrate that it is possible to separate the effect of density from speed of sound mismatch. A scenario where this is used is in the determination of brine/oil separation in reservoirs and it could be useful in the change of morphology or composition of plaque in arteries that becomes dangerously brittle when it hardens with a concomitant increase in the speed of sound. Another scenario where AVA could offer additional information if the two media have the same P-wave velocities and density, but different shear wave (S-wave) velocities, β [13]. Then the reflection at zero angle of incidence will be zero, but as the angle is increased, the polarity and the values of the reflection coefficients become different than zero. This pattern reflects the impact of β mismatch on the coefficients, as is shown in figure 2a. In ultrasonic images taken at an incident angle of zero, the difference in β is not observed. A more realistic case corresponds to two media whose ratio of velocities α and β are both different. For example, in figure 2b, the acoustic impedance is maintained, while varying α and β. The red/blue lines have identical perturbations in α and β, while the black/cyan lines have 20% larger β perturbation. In the study of biological tissues for medical diagnostics, detection of subtle changes in the Poisson's ratio, through variations in S-wave velocities, could aid in the early detection of cancer [11, 14-16].

### 3.2 Wave distortion at an interface

Whenever the angle of incidence is larger than the critical angle for a given medium (either for the P or S-waves), the refracted wave travels parallel to the interface and the reflected waves pick up a phase shift, this is analog to the total internal reflection of light. For sinusoidal waves, this phase shift is not visible in the shape of the wave, but for wavelets it can be detected as distortion. Figures 3a and 3b illustrate this case for a typical wavelet. This scenario could happen when the speed of sound increases sharply across an interface, for example for a wave going from soft tissue to bone. Physical properties, such as densities and sound speeds in tissues utilized in our simulations, were sourced from references [17-20]. S-wave velocities were considered to be 16.9 m/s for muscle and skin, and 1500 m/s for the skull.

### 3.3 Three-layered media

The case of three media was already considered in the original paper by Zoeppritz [1], but only in the simplified possibility that the three densities were identical. In appendix 3 we deduce the eight equations that relate the densities and sound velocities in the three media to the amplitudes in general. One possible application of the three-layered-media model is the determination of the thickness of an intermediate layer using the AVA of the reflected wave. Measuring thickness accurately can be challenging using other methods [8]. The examples in figures 4a and 4b show that the amplitude versus angle of incidence (AVA) that could be used to determine the thickness of the intermediate layer if the other parameters are known. For example, figure 4a shows the effect in the AVA for differences in thicknesses (0.01mm) that are below the wavelength of 0.7mm (for 1 MHz). This does not need to rely on the absolute value of the amplitude, but it could be based on the shape of the function, making it amenable to be treated with modern machine learning algorithms that would classify the shape of the AVA. In principle, the amplitude and the phase shift could be measured as a function of angle for the thickness estimation. In this scenario the technique of lock-in filtering could help to remove noise from the signals and determine the values of amplitude and phase simultaneously. Another possibility is to generate wavelets of short duration and analyze their amplitude and distortion. As an example, figure 5 shows the distorted waves for the case of figure 4a with 1mm thickness and angles of incidence between 0° and 60°.

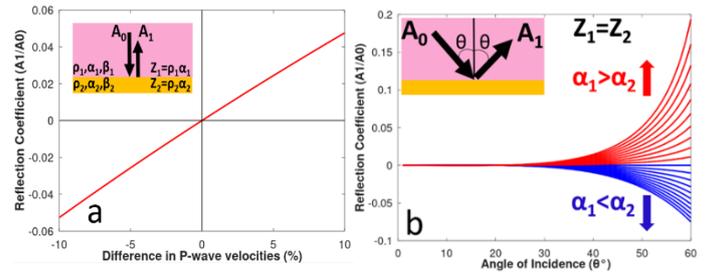

*Figure 1.-(a) Reflection coefficient ($A_1/A_0$) for an angle of incidence zero. Calculated vs. the percentage difference in P-wave velocities between the two media, $100\% \times (\alpha_2 - \alpha_1)/\alpha_1$. In this example $\alpha_1=1500$m/s, $\beta_1=800$m/s, $\alpha_2=1500(1+a)$m/s, $\beta_2=800(1+a)$m/s and $\rho_1=\rho_2=1000$kg/m$^3$, where the perturbation "a" is varied from −10% to +10%. The values of S-wave velocities do not affect the reflected amplitude in this case. (b) Reflection coefficient ($A_1/A_0$) vs angle of incidence. Blue lines are reflection coefficients vs angle of incidence with P-velocities higher in the second media ($\alpha_2 > \alpha_1$) up to 10%, but maintaining the acoustic impedance constant ($z_1 = z_2$). Red lines are for slower velocities ($\alpha_2 < \alpha_1$) up to 10%. In this example $\alpha_1=1500$m/s, $\beta_1=800$m/s, $\rho_1=1000$kg/m$^3$, and $\alpha_2=1500(1+a)$m/s, $\beta_2=800(1+a)$m/s, $\rho_2=1000/(1+a)$kg/m$^3$ with the perturbation "a" between −10% and +10%.*

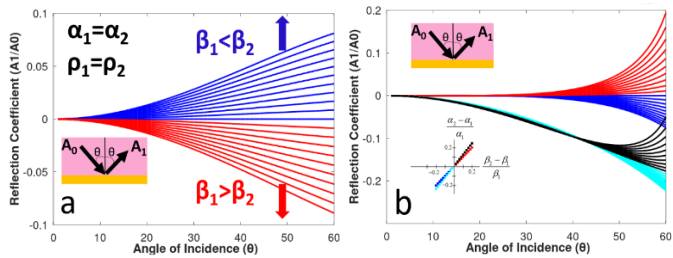

Figure 2.- (a) Effect of S-wave velocities faster (blue) or slower (red). Up to 10%, in the reflection coefficient vs angle of incidence (AVA), while maintaining all other parameters constant. (b) Effect on the AVA if acoustic impedances are equal with different P and S-wave velocities. Notice the large difference between red/blue and black/cyan AVA due to a subtle difference in the ratio $\alpha/\beta$.

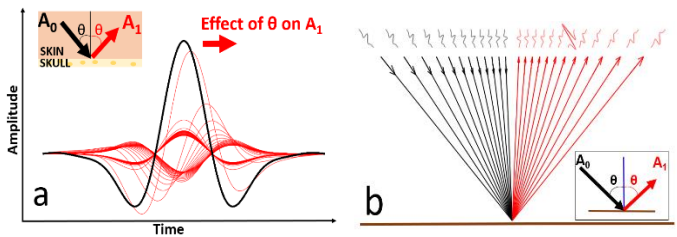

Figure 3.- (a) Family of reflected waveforms (red lines) obtained for angles of incidence between 5° and 60°. This is for reflection of P waves at an interface between skin and skull bone. P-wave velocities are 1615m/s and 2500m/s. Densities are 1090 and 950kg/m$^3$. The reference incident wavelet is shown for comparison (black line). (b) Illustration with the wavelets as a function of angle of incidence. Black are incident and red reflected. Notice the transition after the critical angle of 40°.

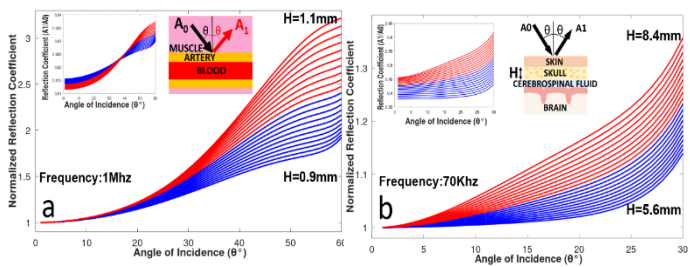

Figure 4.- (a) Normalized reflection coefficient from three-layered media versus angle of incidence. The P-wave velocities are 1547, 1575 and 1584m/s. Densities are 1050, 1060 and 1060 kg/m3. They correspond to muscle, artery wall, and blood respectively. The frequency is 1MHz. The blue family of functions go from 0.90mm to 1.00mm in thickness and the red from 1.01mm to 1.10mm. This demonstrates the possibility of sub-wavelength accuracy in determining the thickness of the intermediate layer. (b) An application of a three-layer reflection in which the second layer (skull) has a large impedance mismatch with the other layers. The frequency is 70kHz in this case. The subtle effect of the second reflection is manifested in the variation with angle of incidence. The inserts show the coefficients before normalizing.

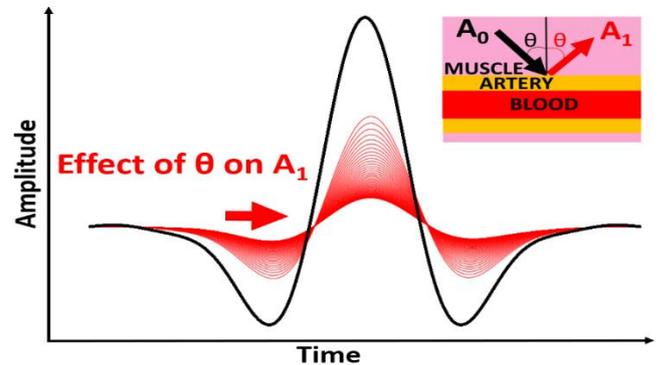

Figure 5.- Family of reflected waveforms (red) obtained for angles of incidence between 0° and 60°.This is for three-layered media as described in Figure 4a with a thickness of 1mm. The reference incident wavelet is shown in black for comparison. This illustrates the possibility of using the distortion of the wave as a tool to measure thickness of an intermediate layer. Reflected intensities were scaled by a factor of 10.

## 4. Conclusions

This work presents potential applications of Z.E.s in medicine, specially thickness determination of artery walls and therefore plaque in the carotid artery. Currently, approximate models based on Z.E.s are successfully utilized in geophysical exploration to study the Earth's subsurface [4]. With the advent of faster and cheaper computers, exact models based on Z.E.s could be applied not only in seismology but also in anatomical exploration such as the prevention of CVAs, which are responsible for paralysis and loss of body functions, ranking as the third cause of death in developed countries [21]. Therefore, we urge the study and employment of Zoeppritz equations along with AVA, the examination of critical angles and waveform distortion as additional tools for the diagnostics and prevention of carotid artery disease. The same methodology could be expanded to measure thickness and density of bones. Overall, angle-dependent reflection coefficients can be integrated into imaging and diagnostic systems and offer data that is not available in other imaging techniques.

**References**

[1] K. Zoeppritz, Uber reflexion und durchgang seismischer wellen durch Unstetigkerlsflaschen: Berlin, Uber Erdbebenwellen VII B, Nachrichten der Koniglichen Gesellschaft der Wissenschaften zu Gottingen, math-phys. Math-Phys., K1. 1919:57-84.

[2] C.G Knott, C.G. Reflexion and refraction of elastic waves, with seismological applications. The London, Edinburgh, and Dublin Philosophical Magazine and Journal of Science. 1899 Jul 1; 48(290):64-97.

[3] J.S. Norris & B.J. Evans, Seismic interface modelling: a physical approach to Zoeppritz Theory. Exploration Geophysics 24(4),1993, 733 - 742

[4] R. T. Shuey (1985). A simplification of the Zoeppritz equations Geophysics, Vol. 50, No. 4, April 1985; P. 609-614

[5] WJ. Ostrander. Plane-wave reflection coefficients for gas sands at nonnormal angles of incidence. Geophysics. 1984 Oct; 49(10):1637-48.

[6] SR. Rutherford, RH. Williams. Amplitude-versus-offset variations in gas sands. Geophysics. 1989 Jun; 54(6):680-8.

[7] J.P. Castagna and M.M. Backus. AVO analysis-tutorial and review, in Castagna, J. and Backus, M.M., eds, Offset-dependent reflectivity. Theory and practice of AVO analysis. Soc.Expl. Geophysics.1993,3-37.

[8] S. James, R. Fedewa, S Lyden, D. Geoffrey. Attenuation compensation comparison for human carotid plaque characterization using spectral analysis of backscattered ultrasound. In 2019 IEEE International Ultrasonics Symposium (IUS) 2019 Oct 6 (pp. 41-44). IEEE

[9] Michael D. Brown et al. Four-dimensional computational ultrasound imaging of brain hemodynamics. Sci Adv. 2024 Jan 19;10(3): eadk7957.doi:10.1126/sciadv.adk7957.

[10] T. Watson, S. El-Jack, JT Stewart, J Ormiston. First-in-human experience using the Volcano VIBE-RX vascular imaging balloon catheter system (Volcano IVUS-guided Balloon Evaluation - New Zealand: VIBE-NZ Study). Euro Intervention. 2013 Sep; 9(5):594-600.

[11] J. Gu et al. Volumetric imaging and morphometric analysis of breast tumor angiogenesis using a new contrast-free ultrasound technique: a feasibility study. Breast Cancer Res 24, 85 (2022).

[12] L.S. Athanasiou, D. I. Fotiadis, L. K. Michalis, 2 - Principles of Coronary Imaging Techniques, Editor(s): Lambros S. Athanasiou, Dimitrios I. Fotiadis, Lampros K. Michalis, Atherosclerotic Plaque Characterization Methods Based on Coronary Imaging, Academic Press, 2017, Pages 23-47, ISBN 9780128047347

[13] K. Helbig and C.S. Mesdag. The potential of shear-wave observations Geophys. Prosp. 30 (1982), 413-431.

[14] J.N. Bruneton, P. Roux, E. Caramella, F. Demard, J. Vallicioni, P. Chauvel. Ear, nose, and throat cancer: ultrasound diagnosis of metastasis to cervical lymph nodes. Radiology. 1984 Sep; 152(3):771-3.

[15] F. Kiessling, J. Bzyl, S. Fokong, M. Siepmann, G. Schmitz, M. Palmowski. Targeted Ultrasound Imaging of Cancer: An Emerging Technology on its Way to Clinics Current Pharmaceutical Design, volume 18, issue 15, pages 2184-2199, year 2012, issn 1381-6128/1873-4286,

[16] J. Künzel, A. Bozzato & S. Strieth. Sonographie in der Nachsorge bei Kopf- und Halskarzinomen. (2017) HNO 65, 939–952

[17] T. Douglas Mast; Empirical relationships between acoustic parameters in human soft tissues. ARLO 1. October 2000; 1 (2): 37-42

[18] P.J. White, G.T. Clement, K. Hynynen. Longitudinal and shear mode ultrasound propagation in human skull bone, Ultrasound in Medicine & Biology, Volume 32, Issue 7, 2006, Pages 1085-1096, ISSN 0301-5629.

[19] MP. Brewin, P.D. Srodon, S.E. Greenwald, M.J. Birch. Carotid atherosclerotic plaque characterisation by measurement of ultrasound sound speed in vitro at high frequency, 20 MHz. Ultrasonics. 2014 Feb;54(2):428-41.

[20] M. Bernabei, SSM. Lee, EJ. Perreault, TG. Sandercock. Shear wave velocity is sensitive to changes in muscle stiffness that occur independently from changes in force. J Appl Physiol(1985). 2020 Jan 1;128(1):8-16

[21] K. Simon, J. Allinson. Deaths related to stroke and cerebrovascular disease, Diagnostic Histopathology, Volume 25, Issue 11, 2019, Pages 444-452, ISSN 1756-2317

# Appendix 1
# On the Reflection and Passage of Seismic
# Waves through Surface Discontinuities

Karl Zoeppritz (deceased).

Presented at the meeting of June 27, 1914, by E. Wiechert.

Professor E. Wiechert addressed the reflection of seismic waves at the Earth's surface in a recently published work (E. Wiechert and K. Zoeppritz: "About Earthquake Waves"). There, the reflection from air and water covers was considered to behave as if they were empty space. While acknowledging this assumption as a rough estimate, which is certainly justified given the inherent inaccuracies still prevalent in seismic measuring methods today, it remains intriguing to develop formulas for the reflection and transmission of elastic waves across the interface of two arbitrary elastic media. These formulas would apply not only to conditions on the Earth's surface, but also to all interfaces within the Earth. Therefore, at the suggestion of Professor Wiechert, I have addressed this more general case and will present some results below.

I am considering the flat interface between two media with densities $\rho_1$ and $\rho_2$, and elastic constants $a_1$, $b_1$ and $a_2$, $b_2$. Where $a$ and $b$ are the speeds of propagation of pure dilation and pure shear waves. As an overview, I will briefly indicate the formulas that connect with Lamé constants $\lambda$ and $\mu$, and Poisson's constant $\sigma$.

$$\frac{\lambda + 2\mu}{\rho} = a^2, \quad \frac{\mu}{\rho} = b^2, \quad \sigma = \frac{a^2 - 2b^2}{2(a^2 - b^2)}$$

Regarding the equations of motion for isotropic, elastic media, we can refer to the previously mentioned work by E. Wiechert. For clarity, I would like to succinctly present it here for one component, using the format commonly employed by English authors[1]:

$$\rho(a^2 - b^2)\frac{\partial \Delta}{\partial x} + \rho b^2 \nabla^2 u = \rho \frac{\partial^2 u}{\partial t^2}$$

Where $\Delta$, the dilation, is $\Delta = \left(\frac{\partial u}{\partial x} + \frac{\partial v}{\partial y} + \frac{\partial w}{\partial z}\right)$

And $\nabla^2$, the Laplace operator, is $\nabla^2 = \frac{\partial^2}{\partial x^2} + \frac{\partial^2}{\partial y^2} + \frac{\partial^2}{\partial z^2}$

While only the two constants $a$ and $b$ occur at the reflection from empty space, we have six in our more general case. Therefore, we should by no means expect such simple formulas as in that special case. I will limit myself to the case of a periodic wave motion. The formulas obtained represent the behavior qualitatively, even in the case of irregular, non-periodic waves. Similar to the reflection on empty space, certain geometric relationships apply between the angles of the incident, reflected and the transmitted wave rays, determined only by the ratios of the speeds. I always consider that the wave comes from Medium 1 onto the flat interface 1, 2 - between Medium 1 and Medium 2. There are two possible cases: the incident wave can be either longitudinal or transverse. In the former case, we are dealing with pure dilation waves. In the latter, conversely, with dilation-free, pure shear waves. I imagine the transversal waves broken down even further into two components, one of which is in the plane of incidence, the other oscillates perpendicular to it. Therefore, we have three cases to investigate, depending on the nature of the incident wave. For the sake of clarity, I will designate the acute angles between ray direction and normal of incidence with two letters. The first of these, *e*, *r*, or *d*, indicates that the ray is incident, reflected or transmitted, the second letter, *l* or *t*, indicates whether the wave is longitudinal or transversal. According to the law of refraction:

$$\sin el : \sin et : \sin rl : \sin rt : \sin dl : \sin dt = a_1 : b_1 : a_1 : b_1 : a_2 : b_2$$

I place the coordinate system in such a way that the *z*-axis points from Medium 1 to Medium 2, in the drawing from bottom to top, the *x*-axis is parallel to the interface in the plane of incidence from left to right. The *y*-axis runs perpendicular to the *x*-axis, extending from the front to the back. I call the displacements *u*, *v*, *w*, accordingly the axes *x*, *y*, *z*, namely $u_1$, $v_1$, $w_1$, for Medium 1, $u_2$, $v_2$, $w_2$, for Medium 2; a horizontal line over the letter means the relevant value for the interface 1, 2.

The following boundary conditions are satisfied at the interface: 1) the sum of the horizontal displacements in Medium 1 is equal to the sum of the horizontal displacements in Medium 2. The same applies to vertical displacements in both media. In formulas:

1) $\sum \bar{u}_1 = \sum \bar{u}_2$    2) $\sum \bar{v}_1 = \sum \bar{v}_2$    3) $\sum \bar{w}_1 = \sum \bar{w}_2$

2) The sum of the pressures at the interface in Medium 1 must be equal to the corresponding of Medium 2. Then they must satisfy the equations:

4) $\sum \rho_1 b_1^2 \left( \dfrac{\partial u_1}{\partial z} + \dfrac{\partial w_1}{\partial x} \right) = \sum \rho_2 b_2^2 \left( \dfrac{\partial u_2}{\partial z} + \dfrac{\partial w_2}{\partial x} \right)$

5) $\sum \rho_1 b_1^2 \left( \dfrac{\partial v_1}{\partial z} + \dfrac{\partial w_1}{\partial y} \right) = \sum \rho_2 b_2^2 \left( \dfrac{\partial v_2}{\partial z} + \dfrac{\partial w_2}{\partial y} \right)$

6) $\sum \left\{ \rho_1 a_1^2 \bar{\Delta}_1 - 2\rho_1 b_1^2 \left( \dfrac{\partial u_1}{\partial x} + \dfrac{\partial v_1}{\partial y} \right) \right\} = \sum \left\{ \rho_2 a_2^2 \bar{\Delta}_2 - 2\rho_2 b_2^2 \left( \dfrac{\partial u_2}{\partial x} + \dfrac{\partial v_2}{\partial y} \right) \right\}$

Where $\Delta = \left( \dfrac{\partial u}{\partial x} + \dfrac{\partial v}{\partial y} + \dfrac{\partial w}{\partial z} \right)$ is the dilation, and $\bar{\Delta}$ is the value of the dilation at the interface.

### 1. Incoming longitudinal wave.

Only shifts in the plane of incidence occur. I denote the displacements for the incident, the reflected and the transmitted longitudinal waves with $\xi_{el}$, $\xi_{rl}$, $\xi_{dl}$ and accordingly for the transverse wave with $\eta_{et}^{(2)}$, $\eta_{rt}$, $\eta_{dt}$ where the directions assumed to be positive are indicated by arrows in the following figure.

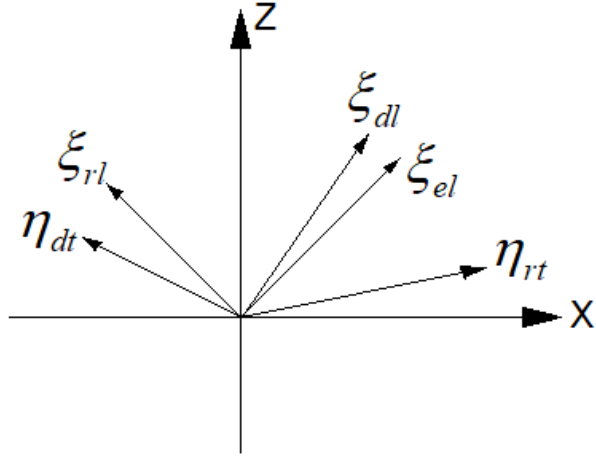

We can make the following assumptions for the plane waves, each of which satisfies the wave equation valid for the Medium.

$$\xi_{el} = R\left\{M_{el} \exp\left[\frac{2\pi i}{T}\left(t - \frac{\sin el\, x + \cos el\, z}{a_1}\right)\right]\right\} \begin{matrix} u = \xi_{el} \sin el \\ v = 0 \\ w = \xi_{el} \cos el \end{matrix}$$

$$\xi_{rl} = R\left\{M_{rl} \exp\left[\frac{2\pi i}{T}\left(t - \frac{\sin el\, x - \cos el\, z}{a_1}\right)\right]\right\} \begin{matrix} u = -\xi_{rl} \sin el \\ v = 0 \\ w = \xi_{rl} \cos el \end{matrix}$$

$$\xi_{dl} = R\left\{M_{dl} \exp\left[\frac{2\pi i}{T}\left(t - \frac{\sin dl\, x + \cos dl\, z}{a_2}\right)\right]\right\} \begin{matrix} u = \xi_{dl} \sin dl \\ v = 0 \\ w = \xi_{dl} \cos dl \end{matrix}$$

$$\eta_{et} = R\left\{0 \exp\left[\frac{2\pi i}{T}\left(t - \frac{\sin et\, x + \cos et\, z}{b_1}\right)\right]\right\} \begin{matrix} u = 0 \\ v = 0 \\ w = 0 \end{matrix}$$

$$\eta_{rt} = R\left\{M_{rt} \exp\left[\frac{2\pi i}{T}\left(t - \frac{\sin et\, x - \cos et\, z}{b_1}\right)\right]\right\} \begin{matrix} u = \eta_{rt} \cos et \\ v = 0 \\ w = \eta_{rt} \sin et \end{matrix}$$

$$\eta_{rdt} = R\left\{M_{dt} \exp\left[\frac{2\pi i}{T}\left(t - \frac{\sin dt\, x + \cos dt\, z}{b_2}\right)\right]\right\} \begin{array}{l} u = -\eta_{dt} \cos dt \\ v = \phantom{-\eta_{dt}} 0 \\ w = \phantom{-}\eta_{dt} \sin dt \end{array}$$

If one incorporates these assumptions into the boundary conditions 1 to 6, omitting conditions 2 and 5 for our case, we end up with four linear and homogeneous equations involving the five quantities $M_{el}$, $M_{rl}$, $M_{dl}$, $M_{rt}$, $M_{dt}$. Assuming that the amplitude $M_{el}$, of the incident wave is known, the other four quantities are then determined by the four boundary conditions. The following matrix is obtained by organizing the ratios of the five quantities $M_{el}$, $M_{rl}$, $M_{dl}$, $M_{rt}$, $M_{dt}$ for the coefficients of the equations:

| $M_{el}$ | $M_{rl}$ | $M_{rt}$ | $M_{dl}$ | $M_{dt}$ |
|---|---|---|---|---|
| $+\sin el$ | $-\sin el$ | $+\cos et$ | $-\sin dl$ | $+\cos dt$ |
| $+\cos el$ | $+\cos el$ | $+\sin et$ | $-\cos dl$ | $-\sin dt$ |
| $-\sin 2el$ | $-\sin 2el$ | $+\dfrac{a_1}{b_1}\cos 2et$ | $+\dfrac{\rho_2}{\rho_1}\left(\dfrac{b_2}{b_1}\right)^2\dfrac{a_1}{a_2}\sin 2dl$ | $-\dfrac{\rho_2}{\rho_1}\left(\dfrac{b_2}{b_1}\right)^2\dfrac{a_1}{a_2}\cos 2dt$ |
| $-\cos 2et$ | $+\cos 2et$ | $+\dfrac{b_1}{a_1}\sin 2et$ | $+\dfrac{\rho_2}{\rho_1}\dfrac{a_2}{a_1}\cos 2dt$ | $+\dfrac{\rho_2}{\rho_1}\dfrac{b_2}{a_1}\sin 2dt$ |

The formulas connecting $M_{rl}$, $M_{rt}$, $M_{dl}$, $M_{dt}$ with $M_{el}$ can be straightforwardly written down based on this matrix, allowing some simplification. However, this simplification is limited since, as noted earlier, we are dealing with six completely different material constants that are independent of each other. The formulas can only be significantly abbreviated if certain simplifying assumptions are made about the nature of these constants. For example, if we set $\rho_2 = 0$, then we have the case of reflection from empty space. Therefore, the top two lines and the last two columns of our matrix disappear, resulting in formulas derived by Prof. Wiechert. By making only $b_2=0$, i.e., that Medium 2 has a stiffness that approaches zero (reflection on liquids or gases), our matrix reads:

| $M_{el}$ | $M_{rl}$ | $M_{rt}$ | $M_{dl}$ | $M_{dt}$ |
|---|---|---|---|---|
| $+\sin el$ | $-\sin el$ | $+\cos et$ | $-\sin dl$ | $1$ |
| $+\cos el$ | $+\cos el$ | $+\sin et$ | $-\cos dl$ | $0$ |
| $-\sin 2el$ | $-\sin 2el$ | $+\dfrac{a_1}{b_1}\cos 2et$ | $0$ | $0$ |
| $-\cos 2et$ | $+\cos 2et$ | $+\dfrac{b_1}{a_1}\sin 2et$ | $+\dfrac{\rho_2}{\rho_1}\dfrac{a_2}{a_1}$ | $0$ |

If one sets the amplitude of the incident wave $M_{el} = 1$, then you can easily derive the formulas from this matrix:

Karl Zoeppritz

$$M_{rl} = -\frac{1-f(1-m)}{1+f(1+m)}$$

$$M_{rt} = \frac{4\dfrac{b_1\rho_1}{a_2\rho_2}\sin el \cos dl \cos 2et}{1+f(1+m)}$$

$$M_{dl} = \frac{2\dfrac{a_1\rho_1}{a_2\rho_2}\cos 2et}{1+f(1+m)}$$

Here $f$ and $m$ represent the following abbreviations:

$$f = \frac{a_1\rho_1}{a_2\rho_2}\frac{\cos dl \cos^2 2et}{\cos el}$$

$$m = \left(\frac{b_1}{a_1}\right)^2 \frac{\sin 2el \tan 2et}{\cos 2et}$$

These formulas are suitable for gaining insights into how a layer of air or water on the surface affects the reflection of longitudinal waves emanating from the depths. The following small table provides the amplitudes $M_{rl}$, and $M_{rt}$, of the reflected longitudinal and transverse waves for some angles of incidence, both in the case of reflection from empty space and in the case of reflection from the seabed. The values of the relevant constants are: 1. For the rock of the earth's surface $\rho=2.7$, $a=7.2 km/s$, $b=4.0 km/s$. 2. For sea water $\rho=1$, $a=1.43 km/s$ and $b=0 km/s$

| Angle of Incidence | $M_{rl}$ Reflection Amplitude | | $M_{rt}$ Reflection Amplitude | |
|---|---|---|---|---|
| | Empty space | Seabed | Empty space | Seabed |
| 0 | 1.00 | 0.87 | 0 | 0 |
| 20 | 0.84 | 0.71 | 0.71 | 0.66 |
| 45 | 0.35 | 0.26 | 1.10 | 1.02 |
| 70 | 0.06 | 0.00 | 0.85 | 0.79 |
| 90 | 0.00 | | 0 | 0 |

As can be seen, the differences between the reflection from the seabed and from empty space are only very slight [3]. With the specified precision of two decimals, you would no longer notice the difference in the case of comparing the reflection from air with that from empty space.

I refrained from applying the obtained formulas to a discontinuous area within the Earth's interior, i.e., to the boundary surface between the rock mantle and the metal core. It is not probable that this transition will change very rapidly. On the contrary, among meteorites, we observe all transition stages from rock meteorites to pure iron meteorites. Therefore, we may reasonably suspect a similar transition, albeit with less certainty, also in the interior of the Earth. On the other hand, such a rapid transition should also be reflected in the seismic velocity curve, especially in the form of a sudden decrease in speed at a certain depth, creating a shadow area for the "first precursors" on the surface of the Earth. However, such evidence has not been demonstrated so far.

### 1. Incoming transverse wave.

a)     Displacement perpendicular to the plane of incidence.

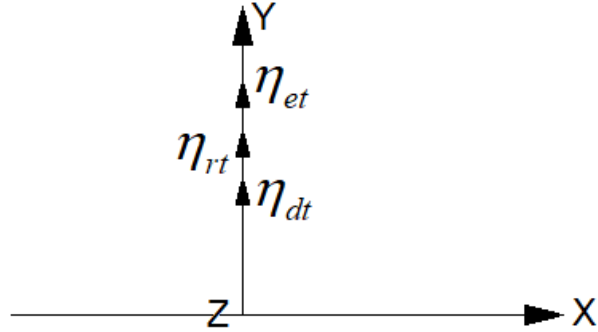

Interface seen from Medium 2.

So, with the transmitted and reflected waves, shifts occur only perpendicular to the plane of incidence, of course. We make the following assumptions to satisfy the wave equation.

$$\eta_{et} = R\left\{M_{et} \exp\left[\frac{2\pi i}{T}\left(t - \frac{x\sin et + z\cos et}{b_1}\right)\right]\right\} \begin{array}{l} u = 0 \\ v = +\eta_{et} \\ w = 0 \end{array}$$

$$\eta_{rt} = R\left\{M_{rt} \exp\left[\frac{2\pi i}{T}\left(t - \frac{x\sin et - z\cos et}{b_1}\right)\right]\right\} \begin{array}{l} u = 0 \\ v = +\eta_{rt} \\ w = 0 \end{array}$$

$$\eta_{dt} = R\left\{M_{dt} \exp\left[\frac{2\pi i}{T}\left(t - \frac{x\sin dt + z\cos dt}{b_2}\right)\right]\right\} \begin{array}{l} u = 0 \\ v = +\eta_{dt} \\ w = 0 \end{array}$$

In this case, the boundary conditions 2 and 5 provide the necessary relationships for determining the ratios $M_{rt}/M_{et}$, $M_{dt}/M_{et}$. The coefficient matrix here is quite straightforward:

| $M_{et}$ | $M_{rt}$ | $M_{dt}$ |
|---|---|---|
| 1 | −1 | $-\dfrac{\rho_2 b_2 \cos dt}{\rho_1 b_1 \cos et}$ |
| 1 | +1 | +1 |

Setting $M_{et} = 1$, produces

$$M_{rt} = +\frac{1-k}{1+k}, \quad M_{dt} = \frac{+2}{1+k}, \quad \text{where } k = \frac{\rho_2}{\rho_1}\frac{b_2}{b_1}\frac{\cos dt}{\cos et}$$

b)   Displacement in the plane of incidence.

Once again, we make specific assumptions to represent plane waves that satisfy the wave equation. The figure indicates the direction in which positive shifts are to be calculated. For transverse waves, the direction of propagation is perpendicular to the direction of displacement, n, rotated clockwise by 90°. The assumptions for the plane waves are as follows:

$$\eta_{et} = R\left\{M_{et}\exp\left[\frac{2\pi i}{T}\left(t-\frac{x\sin et + z\cos et}{b_1}\right)\right]\right\} \quad \begin{array}{l} u = -\eta_{et}\cos et \\ v = 0 \\ w = +\eta_{et}\sin et \end{array}$$

$$\eta_{rt} = R\left\{M_{rt}\exp\left[\frac{2\pi i}{T}\left(t-\frac{x\sin et - z\cos et}{b_1}\right)\right]\right\} \quad \begin{array}{l} u = +\eta_{rt}\cos et \\ v = 0 \\ w = +\eta_{rt}\sin et \end{array}$$

$$\xi_{rl} = R\left\{M_{rl}\exp\left[\frac{2\pi i}{T}\left(t-\frac{x\sin el - z\cos el}{a_1}\right)\right]\right\} \quad \begin{array}{l} u = +\xi_{el}\sin el \\ v = 0 \\ w = +\xi_{el}\cos el \end{array}$$

$$\eta_{dt} = R\left\{M_{dt}\exp\left[\frac{2\pi i}{T}\left(t-\frac{x\sin dt - z\cos dt}{b_2}\right)\right]\right\} \quad \begin{array}{l} u = -\eta_{dt}\cos dt \\ v = 0 \\ w = +\eta_{dt}\sin dt \end{array}$$

$$\xi_{dl} = R\left\{M_{dl}\exp\left[\frac{2\pi i}{T}\left(t-\frac{x\sin dl - z\cos dl}{a_2}\right)\right]\right\} \quad \begin{array}{l} u = +\xi_{dl}\sin dl \\ v = 0 \\ w = +\xi_{dl}\cos dl \end{array}$$

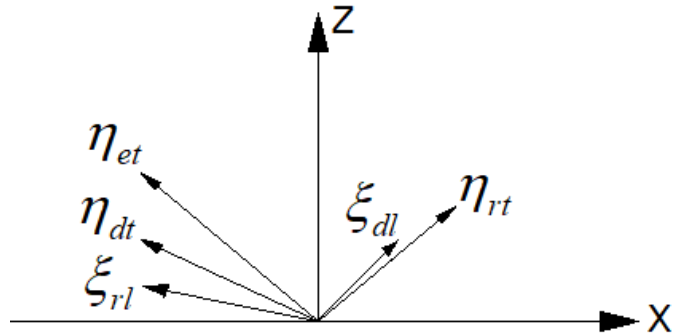

# Karl Zoeppritz

Here again the boundary conditions 1, 3, 4 and 6 are used to determine the amplitude conditions in the following matrix:

$$
\begin{array}{ccccc}
M_{et} & M_{rt} & M_{rl} & M_{dt} & M_{dl} \\
-\cos et & -\cos et & -\sin el & +\cos dt & 1 \\
+\sin et & +\sin et & +\cos el & -\sin dt & 0 \\
+\cos 2et & +\cos 2et & -\dfrac{b_1}{a_1}\sin 2el & -\dfrac{\rho_2}{\rho_1}\dfrac{b_2}{b_1}\cos 2dt & +\dfrac{\rho_2}{\rho_1}\dfrac{b_2}{b_1}\dfrac{b_2}{a_2}\sin 2dt \\
-\sin 2et & +\sin 2et & +\dfrac{a_1}{b_1}\cos 2et & +\dfrac{\rho_2}{\rho_1}\dfrac{b_2}{b_1}\sin 2dt & +\dfrac{\rho_2}{\rho_1}\dfrac{a_2}{b_1}\cos 2dt
\end{array}
$$

Similarly to the incident longitudinal wave, the formulas for the general case are extensive, making it convenient to work directly with the computed matrix. If $b_2 = 0$, meaning Medium 2 has vanishing stiffness, one obtains the following expressions (setting $M_{et} = 1$):

$$M_{rt} = -\frac{1-f(1-m)}{1+f(1+m)}$$

$$M_{rl} = +\frac{\cos dl \sin 4et}{\cos el}\frac{\rho_1 b_1}{\rho_2 a_2}\frac{1}{1+f(1+m)}$$

$$M_{dl} = +\frac{2\sin 2et\, \rho_1 b_1}{\rho_2 a_2}\frac{1}{1+f(1+m)}$$

$$f = \left(\frac{b_1}{a_2}\right)^2 \frac{\rho_1}{\rho_2}\sin 2et \sin 2dl, \quad m = \left(\frac{a_1}{b_1}\right)^2 \frac{\cos 2et}{\sin 2el \tan 2et}$$

Let it be for the case where $b_1 = 0$, the formula for that ratio of horizontal to vertical movement at the limit area is:

$$\frac{\sum \bar{u}}{\sum \bar{w}} = \left(\frac{a_1}{b_1}\right)^2 \frac{\cos 2et}{\sin 2el} - \frac{\rho_2}{\rho_1}\left(\frac{a_2}{b_1}\right)^2 \frac{1}{\sin 2dl}$$

The first term gives the ratio of horizontal motion to vertical movement at the interface, i.e., the tangent of the "apparent" angle of incidence when no Medium 2 is present. The second term accounts for the influence of such a Medium on the apparent angle of incidence.

**Total reflection.**

# Karl Zoeppritz

So far, I have refrained from considering cases where the sine of an angle between the ray and the normal of incidence increases to unity. By analogy with optics or the case of reflection in empty space, we may suspect that when this critical angle is exceeded, in all cases of energy flow, the wave type previously carried has split to the other possible waves. I follow the usual approach in optics, i.e., calculate with imaginary quantities. This leads to imaginary amplitude ratios, allowing us to express the real phase shift of the waves in question compared to the incident one. Simultaneously, it is observed that in all such cases, a boundary wave occurs, the amplitude of which decreases exponentially with distance from the interface and propagates along the interface. No fundamental difficulties arise in any of these cases as long as we focus on considerations limited to periodic waves. However, it is currently not particularly pertinent to formulate formulas for other cases.

If, at sufficiently large angles of incidence, a transverse wave with shifts in the plane of incidence becomes noticeable at the interface, it is possible that both for the reflective as well as for the continuous longitudinal wave, the critical angle has already been exceeded. In such a scenario, it is expected two independent surface waves. However, the energy of the incident transverse wave will be divided between the reflected and the transmitted transverse wave. The formulas for this case are already becoming quite cumbersome. Here I will limit myself to the simpler ones, where the incident transverse wave has displacements perpendicular to the plane. The velocity $b_2$ in Medium 2 is so large that for angles greater than $et_0$:

$$\frac{b_2}{b_1} \sin et > 1, \text{ so } \cos dt = \sqrt{1 - \left(\frac{b_2}{b_1} \sin et\right)^2}$$

becomes imaginary. The quantity $k$ introduced above becomes:

$$k = \frac{\rho_2 b_2}{\rho_1 b_1} \frac{\sqrt{\left(\frac{b_2}{b_1} \sin et\right)^2 - 1}}{\cos et} \sqrt{-1} = i\chi, \text{ where } \chi = \frac{\rho_2 b_2}{\rho_1 b_1} \frac{\sqrt{\left(\frac{b_2}{b_1} \sin et\right)^2 - 1}}{\cos et}$$

$$M_{rt} = \frac{1-i\chi}{1+i\chi} = \frac{1-\chi^2-2i\chi}{1+\chi^2} = \frac{1-\chi^2}{1+\chi^2} + i\frac{-2\chi}{1+\chi^2}$$

$$M_{dt} = \frac{2}{1+i\chi} = \frac{2-2i\chi}{1+\chi^2} = \frac{1}{1+\chi^2} + i\frac{-2\chi}{1+\chi^2}$$

$$\eta_{et} : R = 1\cos\left[\frac{2\pi}{T}\left(t - \frac{x\sin et + z\cos et}{b_1}\right)\right]$$

$$\eta_{rt} : R = \frac{1-\chi^2}{1+\chi^2} \cos\left[\frac{2\pi}{T}\left(t - \frac{x\sin et - z\cos et}{b_1}\right)\right] + \frac{2\chi}{1+\chi^2} \sin\left[\frac{2\pi}{T}\left(t - \frac{x\sin et - z\cos et}{b_1}\right)\right]$$

$$\eta_{dt} : R = \exp\left[-\frac{2\pi}{T}\sqrt{\left(\frac{b_2}{b_1}\sin et\right)^2 - 1}\cdot z \left\{\frac{1-\chi^2}{1+\chi^2}\cos\left[\frac{2\pi}{T}\left(t - \frac{x\sin et}{b_1}\right)\right] + \frac{2\chi}{1+\chi^2}\sin\left[\frac{2\pi}{T}\left(t - \frac{x\sin et}{b_1}\right)\right]\right\}\right]$$

$$\eta_{rt} : R = 1\cdot\cos\left[\frac{2\pi}{T}\left(t - \frac{x\sin et - z\cos et}{b_1}\right) + \alpha\right]$$

$$\tan\alpha = \frac{-2\chi}{1-\chi^2}$$

$$\eta_{dt} : R = \exp\left[-\frac{2\pi}{T}\sqrt{\left(\frac{b_2}{b_1}\sin et\right)^2 - 1}\cdot z \frac{2}{\sqrt{1+\chi^2}}\cos\left\{\frac{2\pi}{T}\left(t - \frac{x\sin et}{b_1}\right) - \beta\right\}\right]$$

$$\tan\beta = -\chi$$

Indeed, these formulas illustrate what is expected:

1. The amplitude of the reflected transverse wave is equal to that of the incident wave, i.e., the energy of the incident wave is totally reflected. The reflected wave undergoes a phase change around the angle.
2. The reflected wave exhibits a phase shift of a certain amount against the incident wave. Instead of a continuous shift, we have a surface wave in the x-direction, decreasing exponentially in the z-direction.

## Magma layers

According to the current state of our knowledge about the interior of the Earth (cf. the relevant explanations in the simultaneously published work "About Seismic Waves" by E. Wiechert and K. Zoeppritz), we must conceive that the Earth offers high resistance to rapidly acting deforming forces, indicating a high degree of rigidity. This is particularly evident in the behavior of shear waves propagating through the Earth's interior. Despite this, it is conceivable, albeit not conclusively, that a layer of reduced stiffness or even a viscous magma layer might be present at a shallow depth. However, such a layer would likely have a minimal impact on the seismic velocity curve. The previously mentioned result regarding the stiffness of the Earth's interior has been derived from this curve. Since, at present, we are unable to obtain definitive answers to this question solely from the seismic velocity curve. It is worth investigating how transverse waves are able to penetrate such a layer. Less suitable for reaching a conclusion in this regard are transverse waves with displacements in the plane of incidence. This is because such waves are always excited when a longitudinal wave encounters a discontinuity surface. The situation is much clearer, however, for transverse waves with vibrations perpendicular to the plane of incidence.

Transverse waves cannot be expelled or absorbed by longitudinal waves and must, therefore, to some extent, preserve their independence as they propagate through the Earth's interior. It has been emphasized that in the "second precursors," vibrations perpendicular to the incidence stand out particularly strongly. This leads us to the question: how can transverse waves perpendicular to the plane of incidence penetrate a thin magma layer? Answering this question requires us to establish what we know about the existence of such a magma layer based on the characteristics of the transverse waves that pass through it.

Let's first consider the case of a liquid, although a very viscous layer, and then briefly touch upon the case of an elastic but less rigid layer. The relationships between the equations in the theory of elasticity and those in the hydrodynamics are crucial here. If we consider the magma layer as a compressible liquid, specifically as an ideal liquid without internal friction, then transverse waves cannot penetrate it. Therefore, this option is excluded. However, the situation is different with viscous liquids. Transverse waves are possible in this case. We can directly apply a solution derived by Stokes (4) for our case of plane waves. The hydrodynamic equations, considering friction, are reduced in the case of flat shear waves to those known from linear heat conduction equations:

$$\frac{\partial v}{\partial t} = \frac{\mu}{\rho}\frac{\partial^2 v}{\partial x^2}$$

where $\mu$ is the coefficient of internal friction, and $\rho$ is the density. If $v$, the transverse displacement, is proportional to $\exp(2i\pi t/T)$, then the solution takes the form:

$$v = Me^{-\gamma x}\cos\left[\frac{2\pi}{T}\left(t - \gamma\frac{T}{2\pi}x\right)\right], \quad \text{where } \gamma = \sqrt{\frac{\pi\rho}{T\mu}}$$

So, the transverse wave propagates, with an exponentially decreasing amplitude, into the liquid. The phase velocity is:

$$b_1 = \frac{2\pi}{T}\frac{1}{\gamma} = \sqrt{\frac{4\pi}{T}\frac{\mu}{\rho}}$$

Thus, it increases with the square root of the friction coefficient and with the frequency $n = 2\pi/T$. Also, the damping travel over the same distance $x$, increases with increasing friction coefficients. For a liquid with the viscosity of water, the propagation speed for $T = 10$ seconds is about 1mm, and for glycerin, it is a few centimeters. The amplitude decreases in water over a path length of 1 cm to about 1/400 of its original value. The same occurs in glycerin over a path length of about 60 cm. Therefore, we would need a viscosity for the magma layer that is a whole order of magnitude greater, if transverse waves are to pass through it.

The coefficient of friction of pitch at 0° was determined by G. H. Darwin (5) in magnitude to $10^{11}$ certainly. With this value, a propagation speed of nearly 3 km/s is obtained for a period of $T=10$s. The amplitude drops to $1/e=0.37$ in about 4 km distance. As can be seen, the rate of propagation is now approaching mean seismic periods of elastic shear wave velocity in solid rock (approximately 4km/s).

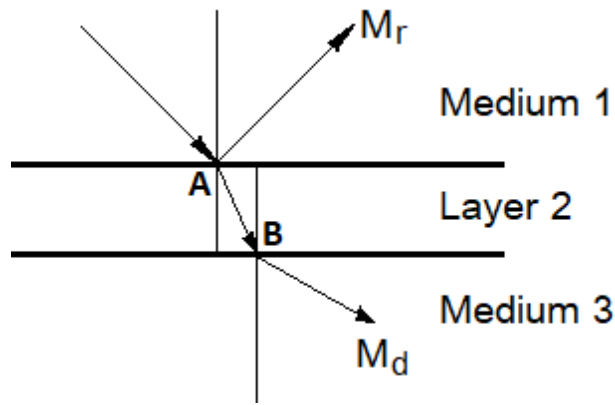

A significant fraction of the wave now penetrates the magma layer, whereas with thin, less viscous magma, almost the entire wave energy would have to be reflected upon hitting the interface, not just due to damping but also because of the significant jump in speed. This difficulty is overcome with an extremely tough medium. Thus, the differences from elastic wave motion are primarily: 1. the dependence of the propagation speed on the oscillation period and 2. the strong damping caused by internal friction. Notably, we must pay attention to the substantial damping even with high toughness. For now, we obtain the result: If a viscous magma layer is present at a certain depth, it is, in any case, only very thin (on the order of perhaps 10 km). Otherwise, transverse waves are unable to penetrate it to any appreciable extent. Let's now, secondarily, attempt to estimate the permeability to transverse waves based on a firm elastic layer but with reduced rigidity. In this case, the influence of internal friction may be much less. Therefore, we neglect it completely and seek the weakening of a transverse wave during passage through a less elastic layer due to reflection alone at the interface. I consider a layer of diminished stiffness inserted between rocks of normal elastic properties. As far as the material constant is concerned, the density differences are, in any case, only very small.

Therefore, $\rho_1 = \rho_2 = \rho_3$. For the calculation of amplitude ratios, we need to apply the formulas established above for the case of an incident wave with displacements perpendicular to the plane of incidence. The strength of the wave reflected backward at A into Medium 1 and the wave penetrating into Medium 3 at B depends, with constant density in all three media, solely on the velocity change when transitioning from 1 to 2 and from 2 to 3. Therefore, I directly measure the reduction in stiffness in layer 2 through the velocity change, denoting it as $b_1 = b_2$. The following small table provides the desired amplitudes $M_r$ and $M_d$ for some incident angles, expressed as a fraction of the amplitude of the wave incident at A. The amplitudes are calculated for the velocity decrease specified in the first row.

The table also allows for the immediate application to friction waves with the same velocity in a viscous fluid. The friction coefficients, corresponding to the respective velocity values for a period of 10 seconds, are provided in the next row. Here, the density is set to 2.7. Finally, the distance is given in another row, to which the amplitude of the transverse wave decreases to $1/e = 0.37$ due to damping in a liquid layer of the specified viscosity.

| Angle of incidence | $b_1:b_2$ = 4:4 | $b_1:b_2$ = 4:3 | $b_1:b_2$ = 4:1 | $b_1:b_2$ = 4:0.4 | $b_1:b_2$ = 4:0.04 | $b_1:b_2$ = 4:0 |
|---|---|---|---|---|---|---|
| | $M_r \mid M_d$ | $M_r \mid M_d$ | $M_r \mid M_d$ | $M_r \mid M_d$ | $M_r \mid M_d$ | $M_r \mid M_d$ |
| 90° | -1 \| 0 | -1 \| 0 | -1 \| 0 | -1 \| 0 | -1 \| 0 | -1 \| 0 |
| 60° | 0 \| 1 | -0.06 \| .99 | 0.34 \| .88 | 0.60 \| .64 | 0.95 \| .098 | 1 \| 0 |
| 30° | 0 \| 1 | 0.11 \| .99 | 0.55 \| .69 | 0.75 \| .44 | 0.97 \| .057 | 1 \| 0 |
| 0° | 0 \| 1 | 0.14 \| .98 | 0.60 \| .65 | 0.82 \| .33 | 0.98 \| .039 | 1 \| 0 |
| Friction coefficient | $3.4 \times 10^{11}$ | $1.9 \times 10^{11}$ | $2.1 \times 10^{10}$ | $2.1 \times 10^8$ | $2.1 \times 10^6$ | 0 |
| Damping length | 6.3 km | 4.7 km | 1.6 km | 0.16 km | 0.016 km | 0 |

The table teaches us how much weakening a transverse wave experiences when passing through a layer of reduced stiffness or through a corresponding viscous layer. In the latter case, there is also the additional weakening due to internal friction, which is so strong that even with not exceptionally high viscosity, it must make the passage of a transverse wave impossible. Moreover, it is crucial to consider that a transverse wave reaching a distant station must traverse the specified layer twice: once during penetration and again upon its return from the Earth's interior. Consequently, the attenuation experienced by the wave as it penetrates to the Earth's surface is compounded.

Summarizing this, if we consider only the result that in the "second precursors," transverse waves perpendicular to the plane of incidence usually emerge quite strongly, it appears unlikely that such a layer is generally present in widespread distribution. E. Wiechert recently emphasized the presence of a layer with increased flexibility to explain transverse surface waves. According to this, an outer solid layer of rock with a thickness of about 25–85 km, at least at the source of earthquakes accompanied by long-period transverse surface waves, would be separated from the rigid Earth's core by a compliant magma layer. Hence, it would be capable of entering free transverse vibrations like a plate bounded on both sides by "free" planes. Based on our formulas, we can now examine whether the boundary of the outer solid crust against the presumed magma layer can indeed be considered "free" when at the same time transverse waves with displacements perpendicular to the plane of incidence are supposed to pass through the layer. Free transverse vibrations of a plate are evidently possible only if the oscillation is almost without weakening reflected back into the plate from both boundary planes of the plate. During each oscillation, a transverse wave, so to speak, falls on the lower boundary at the incident angle. If only a part of the

# Karl Zoeppritz

energy is reflected here, the oscillation process in the plate must run more or less strongly damped depending on the proportion that has penetrated into the underlying magma layer. Referring to our table on p.81 (previous page), we see that even with a velocity decrease in the underlying layer to one-tenth (from 4 km/s to 0.4 km/s), the amplitude of the reflected wave still decreases by about 1/5 with each reflection. Thus, after 3 oscillations or a period of 18 seconds, even before one minute has elapsed, the amplitude of the "free" transverse vibration would have dropped to about half. To explain the emergence of a longer series of transverse surface waves with gradually decreasing amplitude due to free plate vibrations, we would have to assume an even stronger velocity decrease from the solid outer to the compliant inner layer, perhaps from 4 to 0.04 km/s. In this case, however, as our table teaches, the amplitude of the transmitted transverse wave, after passing through the layer once, would decrease to about 4%, and even to less than 2% after passing through twice. The transverse wave, with oscillations perpendicular to the plane of incidence, would likely go undetected due to its weakness. Hence, the presence of intense vibrations perpendicular to the plane of incidence in the "second precursors" appears to contradict the assumption of a flexible layer at a shallow depth beneath the source region.

If we consider the lower stiffness of the layer caused by a softening of the rocks, not only will there be a decrease in velocity for shear waves, but also internal friction will come into play, further unfavorable the picture, as it simultaneously entails additional weakening of the transmitted wave and increased damping of the surface layer wave. Our result would thus be: No generally widespread magma layer. It appears noteworthy to point out that this result is in good agreement with what geology teaches us. For various reasons, modern geology now views the liquid magma reservoirs, whose visible traces confront us in volcanoes, as localized phenomena within the solid Earth's crust and not connected to any potentially universally distributed magma. It would be highly interesting to conduct earthquake observations as close as possible to a large, still active volcano.

In this scenario, it could be possible to observe the damping effect on precursor waves, which could be attributed to the magma at depth exerting influence, particularly if the liquid reservoir is extensive enough. The question about the origin of transverse surface waves remains open in our interpretation. However, it must be said that the observational material in this direction still requires further extensive comparative analysis. In this regard, it would be extremely desirable for the earthquake organization's central offices to release the records of all sufficiently magnifying seismographs suitable for measuring true ground motion—primarily damped instruments or those with undamped apparatus where the natural period exceeds all ground periods—for stronger earthquakes. This should be done in conjunction with a purposeful expansion of the observation network.

Regarding the interpretation of transverse surface waves, it should be noted that the assumption under which Huygens' Principle, applied to earthquake rays, yields a straight seismic ray, namely the assumption of equal velocity in all neighboring points, will not be very extensively fulfilled for surface waves. It can be assumed that differences in the elastic properties of Earth materials, especially in the outer layers of the surface, come into play. This circumstance, along with irregularities in the surface relief, can lead to the vibration direction of surface waves, even if the wave type has a longitudinal character, being strong and deviating from the direction of the epicenter to the earthquake station, changing as the waves propagate. Overall, we are dealing with problems for which the definitive resolution will only come in the future.

Karl Zoeppritz

Gottingen, Geophysical Institute. December 22, 1907.

(1) cf. A. E. H. Lowe, Textbook of Elasticity, German by A. Timpe, 1907.

(2) Doesn't occur in our case.

(3) This prospect uncovered by earthquake research is indeed surprising. Specifically, the distinction in reflection between the seabed and the mainland might not be as insignificant as previously thought. Consequently, it could be possible to determine whether a wave is reflecting from land or from the seabed based on their physical characteristics. This discovery raises the possibility that earthquake monitoring could assist in resolving the longstanding question of delineating water and land boundaries at the poles.

(4) Compare Rayleigh, Theory of sound, first edition p347

(5) Phil. Transactions. 170, S. 1, 1878

(6) E. Wiechert and K. Zoeppritz, 1. c.
   Kgl. Ges. d. Wiss. Nachrichten. Math.-phys. Klasse, 1919, Heft 1

# Karl Zoeppritz
## Appendix 2
## Re-derivation of the Zoeppritz Equations

For P and Sv incident waves at angles $\varphi_0$ and $\psi_0$ with amplitudes $A_0$ and $B_0$ at an interface between two media.

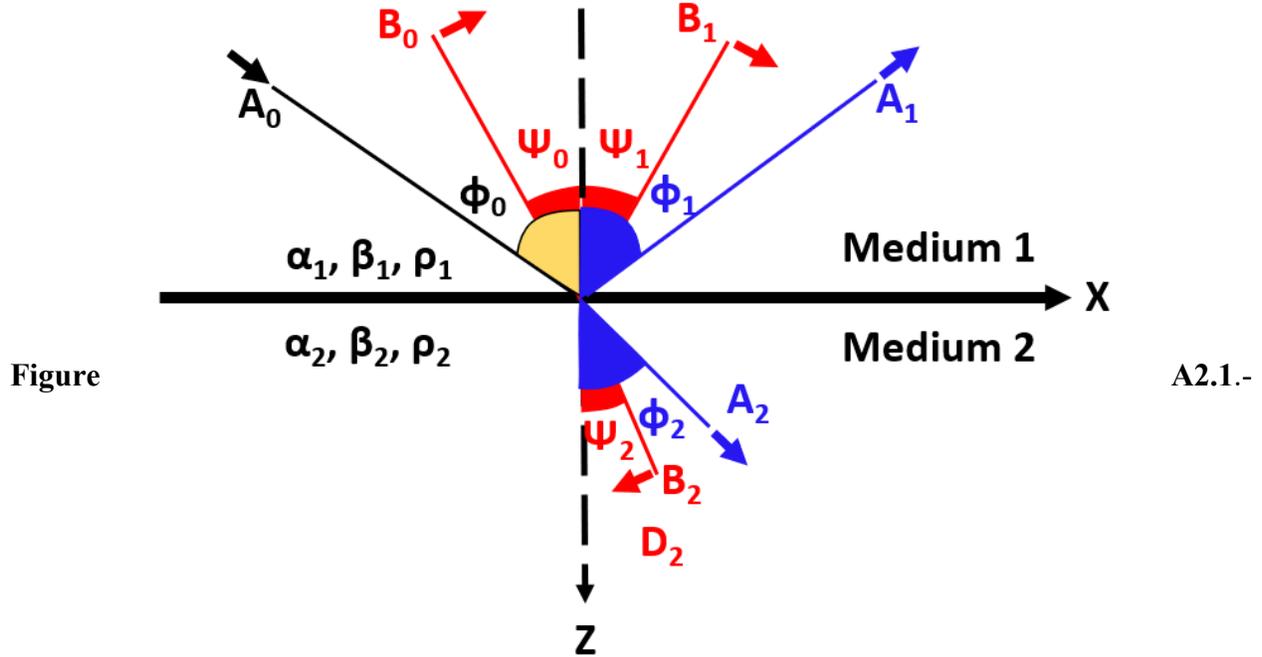

**Figure A2.1.-**

Geometry of P and Sv incident waves on an interface between two media at an angle $\phi_0$, amplitude $A_0$ and $\psi_0$, amplitude $B_0$ respectively. The reflected P-wave at angle $\phi_1 = \phi_0$ has amplitude $A_1$, the reflected converted Sv-wave has amplitude $B_1$ and angle $\psi_1$, the transmitted P-wave has amplitude $A_2$ and angle $\phi_2$, and the transmitted Sv-wave has amplitude $B_2$ and angle $\psi_2$. The arrows indicate the reference for positive polarities (zero phase angle). $\alpha_1$ and $\alpha_2$ are the P-wave velocities in medium 1 and 2, and $\beta_1$ and $\beta_2$ are the corresponding S-wave velocities. $\rho_1$ and $\rho_2$ are the densities of each medium.

As depicted in Figure A2.1, in medium 1 there are four waves, the incident longitudinal one, the incident transversal one, the reflected longitudinal one and a converted reflected transversal wave. The amplitudes are $A_0$, $B_0$, $A_1$ and $B_1$. For medium 2 there are two waves, the transmitted longitudinal one and the transmitted transversal wave, whose amplitudes are $A_2$ and $B_2$. The S-waves in this case are called $S_v$ waves because their displacements are in the plane of incidence, while $S_h$ waves have the displacement orthogonal to the plane of incidence.

The relation between angles can be determined using Snell's law as follows:

$$\frac{\sin\phi_0}{\alpha_1} = \frac{\sin\psi_0}{\beta_1} = \frac{\sin\phi_1}{\alpha_1} = \frac{\sin\phi_2}{\alpha_2} = \frac{\sin\psi_0}{\beta_1} = \frac{\sin\psi_1}{\beta_1} = \frac{\sin\psi_2}{\beta_2}$$

We assume that the amplitudes of the incident waves is 1. That is, $A_0 = 1$ or $B_0 = 1$. Then, we use equations A2.1-8 to determine the other amplitudes. The arrows in the figure indicate the positive directions for the perturbations, which we write below. Here $u$ and $w$ correspond to the $x$ and $z$ directions.

$$u_{A0} = \text{Re}\left\{\exp\left[\omega\left(t - \frac{x\sin\phi_0 + z\cos\phi_0}{\alpha_1}\right)i\right]\right\}\sin\phi_0$$

$$w_{A0} = \text{Re}\left\{\exp\left[\omega\left(t - \frac{x\sin\phi_0 + z\cos\phi_0}{\alpha_1}\right)i\right]\right\}\cos\phi_0$$

$$u_{B0} = \text{Re}\left\{\exp\left[\omega\left(t - \frac{x\sin\psi_0 + z\cos\psi_0}{\beta_1}\right)i\right]\right\}\cos\psi_0$$

$$w_{B0} = -\text{Re}\left\{\exp\left[\omega\left(t - \frac{x\sin\psi_0 + z\cos\psi_0}{\beta_1}\right)i\right]\right\}\sin\psi_0$$

$$u_{A1} = \text{Re}\left\{A_1\exp\left[\omega\left(t - \frac{x\sin\phi_1 - z\cos\phi_1}{\alpha_1}\right)i\right]\right\}\sin\phi_1$$

$$w_{A1} = -\text{Re}\left\{A_1\exp\left[\omega\left(t - \frac{x\sin\phi_1 - z\cos\phi_1}{\alpha_1}\right)i\right]\right\}\cos\phi_1$$

$$u_{B1} = \text{Re}\left\{B_1\exp\left[\omega\left(t - \frac{x\sin\psi_1 - z\cos\psi_1}{\beta_1}\right)i\right]\right\}\cos\psi_1$$

$$w_{B1} = \text{Re}\left\{B_1\exp\left[\omega\left(t - \frac{x\sin\psi_1 - z\cos\psi_1}{\beta_1}\right)i\right]\right\}\sin\psi_1$$

$$u_{A2} = \text{Re}\left\{A_2\exp\left[\omega\left(t - \frac{x\sin\phi_2 + z\cos\phi_2}{\alpha_2}\right)i\right]\right\}\sin\phi_2$$

$$w_{A2} = \text{Re}\left\{A_2\exp\left[\omega\left(t - \frac{x\sin\phi_2 + z\cos\phi_2}{\alpha_2}\right)i\right]\right\}\cos\phi_2$$

$$u_{B2} = -\text{Re}\left\{B_2\exp\left[\omega\left(t - \frac{x\sin\psi_2 + z\cos\psi_2}{\beta_2}\right)i\right]\right\}\cos\psi_2$$

$$w_{B2} = \text{Re}\left\{B_2\exp\left[\omega\left(t - \frac{x\sin\psi_2 + z\cos\psi_2}{\beta_2}\right)i\right]\right\}\sin\psi_2$$

We use complex valued waves to simplify the calculations, with the understanding that the actual displacements are the real components. Next, we look at the conditions at the interface. In the case of a P incident wave, the subscript *inc* refers to $A_0$. For a S$_V$ wave, *inc* refers to $B_0$. The sum of the displacements at the interface ($z = 0$) must be the same.

$$w_{inc} + w_{A1} + w_{B1}\big|_{z=0} = w_{A2} + w_{B2}\big|_{z=0} \ldots \text{(A2.1)}$$

$$u_{inc} + u_{A1} + u_{B1}\big|_{z=0} = u_{A2} + u_{B2}\big|_{z=0} \ldots \text{(A2.2)}$$

The third condition is that the pressure must be the same at the interface (z=0), so:

$$\sum\left\{\rho_1\alpha_1^2\bar{\Delta}_1-2\rho_1\beta_1^2\left(\frac{\partial u_1}{\partial x}+\frac{\partial v_1}{\partial y}\right)\right\}=\sum\left\{\rho_2\alpha_2^2\bar{\Delta}_2-2\rho_2\beta_2^2\left(\frac{\partial u_2}{\partial x}+\frac{\partial v_2}{\partial y}\right)\right\}\ldots \text{(A2.3)}$$

Where $\Delta$ is the dilation, $\Delta=\left(\frac{\partial u}{\partial x}+\frac{\partial v}{\partial y}+\frac{\partial w}{\partial z}\right)$, $\bar{\Delta}$ is the value of the dilation at the interface between medium 1 and medium 2.

The fourth condition is that the shear strain must be the same at z=0, so

$$\sum\rho_1\beta_1^2\left(\frac{\partial u_1}{\partial z}+\frac{\partial w_1}{\partial x}\right)=\sum\rho_2\beta_2^2\left(\frac{\partial u_2}{\partial z}+\frac{\partial w_2}{\partial x}\right)\ldots \text{(A2.4)}$$

### Equations for an incident P wave ($A_0 =1$ and $B_0 =0$)
Applying equations 1-8, we obtain:

$$\cos\phi_1 A_1 - \sin\psi_1 B_1 + \cos\phi_2 A_2 + \sin\psi_2 B_2 = \cos\phi_0 \ldots \text{(A2.5)}$$

$$-\sin\phi_1 A_1 - \cos\psi_1 B_1 + \sin\phi_2 A_2 - \cos\psi_2 B_2 = \sin\phi_0 \ldots \text{(A2.6)}$$

$$-\cos 2\psi_1 A_1 + \frac{\beta_1}{\alpha_1}\sin 2\psi_1 B_1 + \frac{\rho_2}{\rho_1}\frac{\alpha_2}{\alpha_1}\cos 2\psi_2 A_2 + \frac{\rho_2}{\rho_1}\frac{\beta_2}{\alpha_1}\sin 2\psi_2 B_2 = \cos 2\psi_0 \ldots \text{(A2.7)}$$

$$\sin 2\phi_1 A_1 + \frac{\alpha_1}{\beta_1}\cos 2\psi_1 B_1 + \frac{\rho_2\alpha_1\beta_2^2}{\rho_1\alpha_2\beta_1^2}\sin 2\phi_2 A_2 - \frac{\rho_2\alpha_1\beta_2}{\rho_1\beta_1^2}\cos 2\psi_2 B_2 = \sin 2\phi_0 \ldots \text{(A2.8)}$$

These equations in matrix form for an incident P-wave are:

$$\begin{bmatrix} \cos\phi_1 & -\sin\psi_1 & \cos\phi_2 & \sin\psi_2 \\ -\sin\phi_1 & -\cos\psi_1 & \sin\phi_2 & -\cos\psi_2 \\ -\cos 2\psi_1 & \frac{\beta_1}{\alpha_1}\sin 2\psi_1 & \frac{\rho_2\alpha_2}{\rho_1\alpha_1}\cos 2\psi_2 & \frac{\rho_2\beta_2}{\rho_1\alpha_1}\sin 2\psi_2 \\ \sin 2\phi_1 & \frac{\alpha_1}{\beta_1}\cos 2\psi_1 & \frac{\rho_2\alpha_1\beta_2^2}{\rho_1\alpha_2\beta_1^2}\sin 2\phi_2 & -\frac{\rho_2\alpha_1\beta_2}{\rho_1\beta_1^2}\cos 2\psi_2 \end{bmatrix} \begin{bmatrix} A_1 \\ B_1 \\ A_2 \\ B_2 \end{bmatrix} = \begin{bmatrix} \cos\phi_0 \\ \sin\phi_0 \\ \cos 2\psi_0 \\ \sin 2\phi_0 \end{bmatrix}$$

### Equations for an incident $S_V$ wave ($B_0 =1$ and $A_0 =0$)
Applying equations 1-8, we obtain:

$$\cos\phi_1 A_1 - \sin\psi_1 B_1 + \cos\phi_2 A_2 + \sin\psi_2 B_2 = -\sin\psi_0 \ldots \text{(A2.9)}$$

$$-\sin\phi_1 A_1 - \cos\psi_1 B_1 + \sin\phi_2 A_2 - \cos\psi_2 B_2 = \cos\psi_0 \ldots \text{(A2.10)}$$

$$-\cos 2\psi_1 A_1 + \frac{\beta_1}{\alpha_1}\sin 2\psi_1 B_1 + \frac{\rho_2}{\rho_1}\frac{\alpha_2}{\alpha_1}\cos 2\psi_2 A_2 + \frac{\rho_2}{\rho_1}\frac{\beta_2}{\alpha_1}\sin 2\psi_2 B_2 = -\frac{\beta_1}{\alpha_1}\sin 2\psi_0 \ldots \text{(A2.11)}$$

$$\sin 2\phi_1 A_1 + \frac{\alpha_1}{\beta_1}\cos 2\psi_1 B_1 + \frac{\alpha_1\beta_2^2\rho_2}{\alpha_2\beta_1^2\rho_1}\sin 2\phi_2 A_2 - \frac{\rho_2\alpha_1\beta_2}{\rho_1\beta_1^2}\cos 2\psi_2 B_2 = \frac{\alpha_1}{\beta_1}\cos 2\psi_0 \ldots \text{(A2.12)}$$

In matrix form for an incident $S_V$-wave:

$$\begin{bmatrix} \cos\phi_1 & -\sin\psi_1 & \cos\phi_2 & \sin\psi_2 \\ -\sin\phi_1 & -\cos\psi_1 & \sin\phi_2 & -\cos\psi_2 \\ -\cos 2\psi_1 & \dfrac{\beta_1}{\alpha_1}\sin 2\psi_1 & \dfrac{\rho_2\alpha_2}{\rho_1\alpha_1}\cos 2\psi_2 & \dfrac{\rho_2\beta_2}{\rho_1\alpha_1}\sin 2\psi_2 \\ \sin 2\phi_1 & \dfrac{\alpha_1}{\beta_1}\cos 2\psi_1 & \dfrac{\rho_2\alpha_1\beta_2^{\,2}}{\rho_1\alpha_2\beta_1^{\,2}}\sin 2\phi_2 & -\dfrac{\rho_2\alpha_1\beta_2}{\rho_1\beta_1^{\,2}}\cos 2\psi_2 \end{bmatrix} \begin{bmatrix} A_1 \\ B_1 \\ A_2 \\ B_2 \end{bmatrix} = \begin{bmatrix} -\sin\psi_0 \\ \cos\psi_0 \\ -\dfrac{\beta_1}{\alpha_1}\sin 2\psi_0 \\ \dfrac{\alpha_1}{\beta_1}\cos 2\psi_0 \end{bmatrix}$$

**Incident $S_h$ wave**

Finally, we consider an incident transversal wave with its displacement normal to the plane of incidence, also known as an $S_h$-wave.

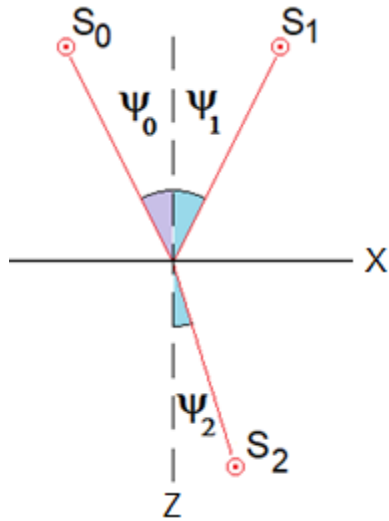

**Figure A2.2.**- Geometry of an $S_h$-wave incident on an interface between two media at an angle $\psi_0$ and amplitude $S_0$. The reflected $S_h$-wave has amplitude $S_1$ and angle $\psi_1 = \psi_0$, the transmitted S-wave has amplitude $S_2$ and angle $\psi_2$. The arrow tips indicate the reference for positive polarities (zero phase angle) out of the figure in the positive y-direction.

As depicted in Figure A2.2, in medium 1 there are two waves, the incident transversal one and a reflected transversal wave. The amplitudes are $S_0$ and $S_1$. For medium 2 there is only one wave, the transversal one, whose amplitude is $S_2$. The S-waves in this case are called $S_h$ because their displacements are orthogonal to the plane of incidence, while $S_V$ waves have the displacement in the plane of incidence.

The relation between angles can be determined using Snell's law as follows:

$$\frac{\sin\psi_0}{\beta_1} = \frac{\sin\psi_1}{\beta_1} = \frac{\sin\psi_2}{\beta_2}$$

Where $\beta_1$ and $\beta_2$ are the S-wave velocities in media 1 and 2.

We let $S_0 = 1$ and use the equations to determine the other amplitudes. The arrows tips in the figure indicate the positive directions for the perturbations, which we write below. Here $v$ is the displacement in the $y$ direction.

$$v_{S0} = \text{Re}\left\{\exp\left[\omega\left(t - \frac{x\sin\psi_0 + z\cos\psi_0}{\beta_1}\right)i\right]\right\}$$

$$v_{S1} = \text{Re}\left\{S_1 \exp\left[\omega\left(t - \frac{x\sin\psi_1 - z\cos\psi_1}{\beta_1}\right)i\right]\right\}$$

$$v_{S2} = \text{Re}\left\{S_2 \exp\left[\omega\left(t - \frac{x\sin\psi_2 + z\cos\psi_2}{\beta_2}\right)i\right]\right\}$$

Similarly to the cases of P and $S_v$ waves, we use complex valued waves with the actual displacements being the real components. Next, we look at the conditions at the interface. The first condition is that the sum of the displacements must be the same.

$$v_{S0} + v_{S1}\big|_{z=0} = v_{S2}\big|_{z=0}$$

$$S_1 - S_2 = -1 \ldots \text{(A2.13)}$$

The other condition is that the shear strain must be the same at z=0, so

$$\sum \rho_1 \beta_1^2 \left(\frac{\partial v_1}{\partial z}\right) = \sum \rho_2 \beta_2^2 \left(\frac{\partial v_2}{\partial z}\right)$$

$$\rho_1 \beta_1 \cos\psi_1 S_1 + \rho_2 \beta_2 \cos\psi_2 S_2 = \rho_1 \beta_1 \cos\psi_0 \ldots \text{(A2.14)}$$

In matrix form

$$\begin{bmatrix} 1 & -1 \\ \rho_1 \beta_1 \cos\psi_1 & \rho_2 \beta_2 \cos\psi_2 \end{bmatrix} \begin{bmatrix} S_1 \\ S_2 \end{bmatrix} = \begin{bmatrix} -1 \\ \rho_1 \beta_1 \cos\psi_0 \end{bmatrix}$$

Karl Zoeppritz
# Appendix 3
# Equations for two interfaces between three different media

For P and S$_V$ incident waves at angles $\phi_0$ and $\psi_0$ with amplitudes A$_0$ and B$_0$.

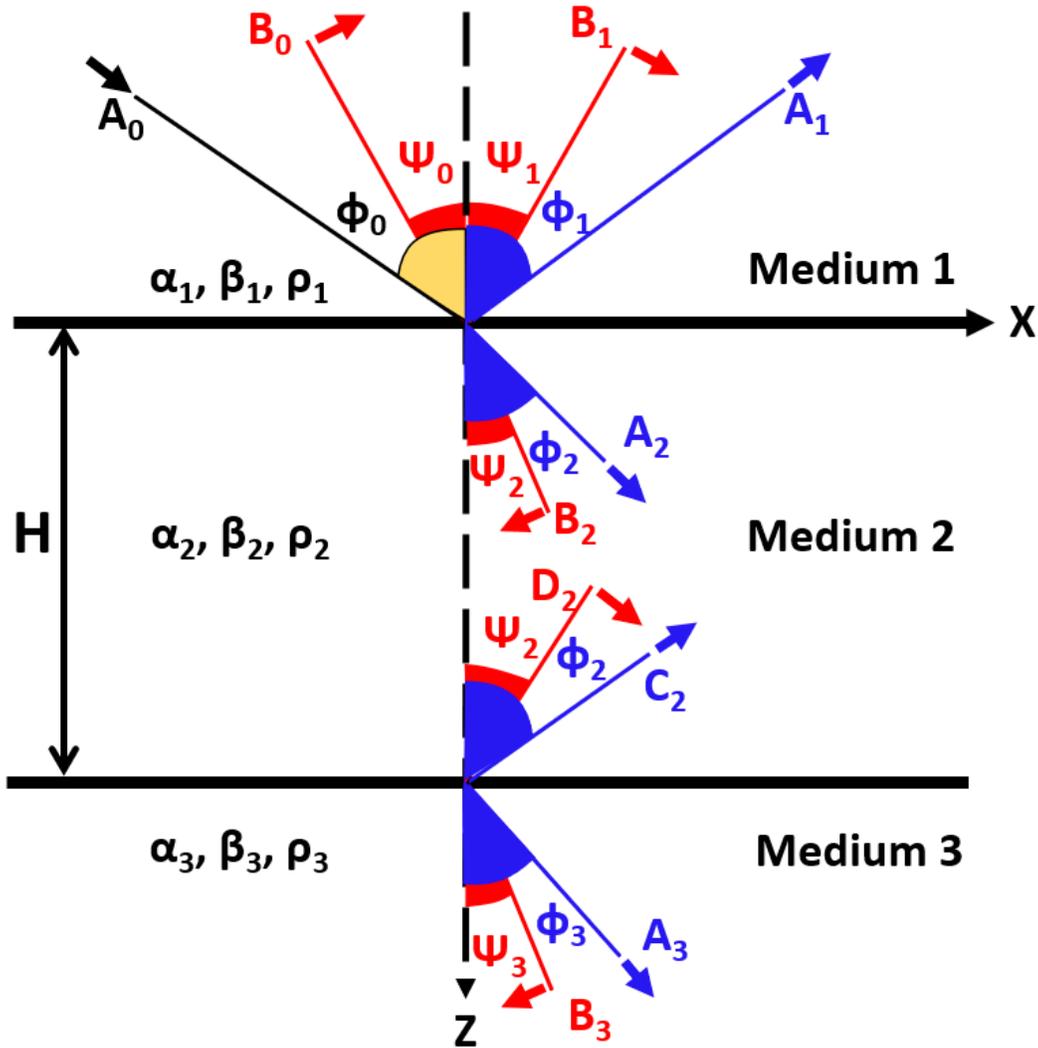

**Figure A3.1** Three layers

Snell's law:
$$\frac{\sin\phi_0}{\alpha_1} = \frac{\sin\phi_1}{\alpha_1} = \frac{\sin\phi_2}{\alpha_2} = \frac{\sin\phi_3}{\alpha_3} = \frac{\sin\psi_0}{\beta_1} = \frac{\sin\psi_1}{\beta_1} = \frac{\sin\psi_2}{\beta_2} = \frac{\sin\psi_3}{\beta_3} = p$$

In medium 1 there are three waves, the incident longitudinal (A$_0$) or the incident transversal wave (B$_0$), the reflected longitudinal one (A$_1$) and a converted reflected transversal wave (B$_1$). For medium 2, there are four waves, the transmitted longitudinal, a converted transmitted transversal wave and correspondingly two reflected waves. The amplitudes are A$_2$, B$_2$ for the waves going downwards and C$_2$, D$_2$ for the reflected ones. Finally, in medium 3 we label the waves A$_3$ and B$_3$ for the longitudinal and transversal ones respectively. We assume that the amplitudes of the incident waves is 1 when present. That is, $A_0 = 1$ or $B_0 = 1$. Then, we use the equations 1-8 to determine the other amplitudes. The arrows in the figure indicate the positive directions for the perturbations, which we write below. Here $u$, $v$ and $w$ correspond to the $x$, $y$ and $z$ directions.

$$u_{A0} = \text{Re}\left\{\exp\left[\omega\left(t - \frac{x\sin\phi_0 + z\cos\phi_0}{\alpha_1}\right)i\right]\right\}\sin\phi_0$$

$$w_{A0} = \text{Re}\left\{\exp\left[\omega\left(t - \frac{x\sin\phi_0 + z\cos\phi_0}{\alpha_1}\right)i\right]\right\}\cos\phi_0$$

$$u_{B0} = \text{Re}\left\{\exp\left[\omega\left(t - \frac{x\sin\psi_0 + z\cos\psi_0}{\beta_1}\right)i\right]\right\}\cos\psi_0$$

$$w_{B0} = -\text{Re}\left\{\exp\left[\omega\left(t - \frac{x\sin\psi_0 + z\cos\psi_0}{\beta_1}\right)i\right]\right\}\sin\psi_0$$

$$w_{A1} = -\text{Re}\left\{A_1\exp\left[\omega\left(t - \frac{x\sin\phi_1 - z\cos\phi_1}{\alpha_1}\right)i\right]\right\}\cos\phi_1$$

$$u_{B1} = \text{Re}\left\{B_1\exp\left[\omega\left(t - \frac{x\sin\psi_1 - z\cos\psi_1}{\beta_1}\right)i\right]\right\}\cos\psi_1$$

$$w_{B1} = \text{Re}\left\{B_1\exp\left[\omega\left(t - \frac{x\sin\psi_1 - z\cos\psi_1}{\beta_1}\right)i\right]\right\}\sin\psi_1$$

$$u_{A2} = \text{Re}\left\{A_2\exp\left[\omega\left(t - \frac{x\sin\phi_2 + z\cos\phi_2}{\alpha_2}\right)i\right]\right\}\sin\phi_2$$

$$w_{A2} = \text{Re}\left\{A_2\exp\left[\omega\left(t - \frac{x\sin\phi_2 + z\cos\phi_2}{\alpha_2}\right)i\right]\right\}\cos\phi_2$$

$$u_{B2} = -\text{Re}\left\{B_2\exp\left[\omega\left(t - \frac{x\sin\psi_2 + z\cos\psi_2}{\beta_2}\right)i\right]\right\}\cos\psi_2$$

$$w_{B2} = \text{Re}\left\{B_2\exp\left[\omega\left(t - \frac{x\sin\psi_2 + z\cos\psi_2}{\beta_2}\right)i\right]\right\}\sin\psi_2$$

$$u_{C2} = \text{Re}\left\{C_2\exp\left[\omega\left(t - \frac{x\sin\phi_2 - (z-H)\cos\phi_2}{\alpha_2}\right)i\right]\right\}\sin\phi_2$$

$$w_{C2} = -\text{Re}\left\{C_2\exp\left[\omega\left(t - \frac{x\sin\phi_2 - (z-H)\cos\phi_2}{\alpha_2}\right)i\right]\right\}\cos\phi_2$$

$$u_{D2} = \text{Re}\left\{D_2\exp\left[\omega\left(t - \frac{x\sin\psi_2 - (z-H)\cos\psi_2}{\beta_2}\right)i\right]\right\}\cos\psi_2$$

Karl Zoeppritz

$$w_{D2} = \text{Re}\left\{D_2 \exp\left[\omega\left(t - \frac{x\sin\psi_2 - (z-H)\cos\psi_2}{\beta_2}\right)i\right]\right\}\sin\psi_2$$

$$u_{A3} = \text{Re}\left\{A_3 \exp\left[\omega\left(t - \frac{x\sin\phi_3 + (z-H)\cos\phi_3}{\alpha_3}\right)i\right]\right\}\sin\phi_3$$

$$w_{A3} = \text{Re}\left\{A_3 \exp\left[\omega\left(t - \frac{x\sin\phi_3 + (z-H)\cos\phi_3}{\alpha_3}\right)i\right]\right\}\cos\phi_3$$

$$u_{B3} = -\text{Re}\left\{B_3 \exp\left[\omega\left(t - \frac{x\sin\psi_3 + (z-H)\cos\psi_3}{\beta_3}\right)i\right]\right\}\cos\psi_3$$

$$w_{B3} = \text{Re}\left\{B_3 \exp\left[\omega\left(t - \frac{x\sin\psi_3 + (z-H)\cos\psi_3}{\beta_3}\right)i\right]\right\}\sin\psi_3$$

Next, we look at the conditions at the border. In the case of a P incident wave, the subscript *inc* refers to $A_0$. For a $S_V$ wave, *inc* refers to $B_0$. The sum of the displacements at the interface ($z = 0$) must be the same.

For displacement in the z-direction.

$$w_{inc} + w_{A1} + w_{B1}\big|_{z=0} = w_{A2} + w_{B2} + w_{C2} + w_{D2}\big|_{z=0} \quad \ldots \text{(A3.1)}$$

For displacement in the x-direction.

$$u_{inc} + u_{A1} + u_{B1}\big|_{z=0} = u_{A2} + u_{B2} + u_{C2} + u_{D2}\big|_{z=0} \quad \ldots \text{(A3.2)}$$

Similarly, the sum of the displacements at the interface ($z = H$) must be the same.

$$w_{A2} + w_{B2} + w_{C2} + w_{D2}\big|_{z=H} = w_{A3} + w_{B3}\big|_{z=H} \quad \ldots \text{(A3.3)}$$

$$u_{A2} + u_{B2} + u_{C2} + u_{D2}\big|_{z=H} = u_{A3} + u_{B3}\big|_{z=H} \quad \ldots \text{(A3.4)}$$

The next conditions are that the pressure has to be the same at the interfaces, so:
At z = 0

$$\sum\left\{\rho_1\alpha_1^2\overline{\Delta}_1 - 2\rho_1\beta_1^2\left(\frac{\partial u_1}{\partial x} + \frac{\partial v_1}{\partial y}\right)\right\} = \sum\left\{\rho_2\alpha_2^2\overline{\Delta}_2 - 2\rho_2\beta_2^2\left(\frac{\partial u_2}{\partial x} + \frac{\partial v_2}{\partial y}\right)\right\} \quad \ldots \text{(A3.5)}$$

And at z= H

$$\sum\left\{\rho_2\alpha_2^2\overline{\Delta}_2 - 2\rho_2\beta_2^2\left(\frac{\partial u_2}{\partial x} + \frac{\partial v_2}{\partial y}\right)\right\} = \sum\left\{\rho_3\alpha_3^2\overline{\Delta}_3 - 2\rho_3\beta_3^2\left(\frac{\partial u_3}{\partial x} + \frac{\partial v_3}{\partial y}\right)\right\} \quad \ldots \text{(A3.6)}$$

Where $\Delta$ is the dilation, $\Delta = \left(\frac{\partial u}{\partial x} + \frac{\partial v}{\partial y} + \frac{\partial w}{\partial z}\right)$ and $\overline{\Delta}$ is the value at interfaces.

The final conditions are that the shear strains have to be the same at z=0

$$\sum \rho_1\beta_1^2\left(\frac{\partial u_1}{\partial z} + \frac{\partial w_1}{\partial x}\right) = \sum \rho_2\beta_2^2\left(\frac{\partial u_2}{\partial z} + \frac{\partial w_2}{\partial x}\right) \quad \ldots \text{(A3.7)}$$

and at z=H

$$\sum \rho_2 \beta_2{}^2 \left( \frac{\partial u_2}{\partial z} + \frac{\partial w_2}{\partial x} \right) = \sum \rho_3 \beta_3{}^2 \left( \frac{\partial u_3}{\partial z} + \frac{\partial w_3}{\partial x} \right) \quad \ldots \text{(A3.8)}$$

**Equations for an incident P wave ($A_0 = 1$ and $B_0 = 0$)**

Applying equations 1-8, we obtain:

$$\cos\phi_1 A_1 - \sin\psi_1 B_1 + \cos\phi_2 A_2 + \sin\psi_2 B_2 - \cos\phi_2 F_l C_2 + \sin\psi_2 F_t D_2 = \cos\phi_0 \quad \ldots \text{(A3.9)}$$

where

$$F_l = \exp\left( -\frac{H\omega \cos\phi_2}{\alpha_2} i \right)$$

$$F_t = \exp\left( -\frac{H\omega \cos\psi_2}{\beta_2} i \right)$$

$$-\sin\phi_1 A_1 - \cos\psi_1 B_1 + \sin\phi_2 A_2 - \cos\psi_2 B_2 + \sin\phi_2 F_l C_2 + \cos\psi_2 F_t D_2 = \sin\phi_0 \quad \ldots \text{(A3.10)}$$

$$\cos\phi_2 F_l A_2 + \sin\psi_2 F_t B_2 - \cos\phi_2 C_2 + \sin\psi_2 D_2 - \cos\phi_3 A_3 - \sin\psi_3 B_3 = 0 \quad \ldots \text{(A3.11)}$$

$$\sin\phi_2 F_l A_2 - \cos\psi_2 F_t B_2 + \sin\phi_2 C_2 + \cos\psi_2 D_2 - \sin\phi_3 A_3 + \cos\psi_3 B_3 = 0 \quad \ldots \text{(A3.12)}$$

At z=0

$$-\cos 2\psi_1 A_1 + \frac{\beta_1}{\alpha_1} \sin 2\psi_1 B_1 + \frac{\rho_2 \alpha_2}{\rho_1 \alpha_1} \cos 2\psi_2 A_2 + \frac{\rho_2 \beta_2}{\rho_1 \alpha_1} \sin 2\psi_2 B_2$$

$$+ \frac{\rho_2 \alpha_2}{\rho_1 \alpha_1} \cos 2\psi_2 F_l C_2 - \frac{\rho_2 \beta_2}{\rho_1 \alpha_1} \sin 2\psi_2 F_t D_2 = \cos 2\psi_1 \quad \ldots \text{(A3.13)}$$

And at z= H

$$-\cos 2\psi_2 F_l A_2 - \frac{\beta_2}{\alpha_2} \sin 2\psi_2 F_t B_2 - \cos 2\psi_2 C_2 + \frac{\beta_2}{\alpha_2} \sin 2\psi_2 D_2$$

$$+ \frac{\alpha_3 \rho_3}{\alpha_2 \rho_2} \cos 2\psi_3 A_3 + \frac{\beta_3 \rho_3}{\alpha_2 \rho_2} \sin 2\psi_3 B_3 = 0 \quad \ldots \text{(A3.14)}$$

$$\sin 2\phi_1 A_1 + \frac{\alpha_1}{\beta_1} \cos 2\psi_1 B_1 + \frac{\rho_2 \alpha_1 \beta_2{}^2}{\rho_1 \alpha_2 \beta_1{}^2} \sin 2\phi_2 A_2 - \frac{\rho_2 \alpha_1 \beta_2}{\rho_1 \beta_1{}^2} \cos 2\psi_2 B_2$$

$$- \frac{\rho_2 \alpha_1 \beta_2{}^2}{\rho_1 \alpha_2 \beta_1{}^2} \sin 2\phi_2 F_l C_2 - \frac{\rho_2 \alpha_1 \beta_2}{\rho_1 \beta_1{}^2} \cos 2\psi_2 F_t D_2 = \sin 2\phi_0 \quad \ldots \text{(A3.15)}$$

$$-\sin 2\phi_2 F_l A_2 + \frac{\alpha_2}{\beta_2} \cos 2\psi_2 F_t B_2 + \sin 2\phi_2 C_2 + \frac{\alpha_2}{\beta_2} \cos 2\psi_2 D_2$$

$$\frac{\rho_3 \alpha_2 \beta_3{}^2}{\rho_2 \alpha_3 \beta_2{}^2} \sin 2\phi_3 A_3 - \frac{\rho_3 \alpha_2 \beta_3}{\rho_2 \beta_2{}^2} \cos 2\psi_3 B_3 = 0 \quad \ldots \text{(A3.16)}$$

These equations give rise to matrix form for an incident P-wave

$$\begin{bmatrix} \cos\phi_1 & -\sin\psi_1 & \cos\phi_2 & \sin\psi_2 & -\cos\phi_2 F_l & \sin\psi_2 F_t & 0 & 0 \\ -\sin\phi_1 & -\cos\psi_1 & \sin\phi_2 & -\cos\psi_2 & \sin\phi_2 F_l & \cos\psi_2 F_t & 0 & 0 \\ 0 & 0 & \cos\phi_2 F_l & \sin\psi_2 F_t & -\cos\phi_2 & \sin\psi_2 & -\cos\phi_3 & -\sin\psi_3 \\ 0 & 0 & \sin\phi_2 F_l & -\cos\psi_2 F_t & \sin\phi_2 & \cos\psi_2 & -\sin\phi_3 & \cos\psi_3 \\ -\cos 2\psi_1 & \frac{\beta_1}{\alpha_1}\sin 2\psi_1 & \frac{\rho_2\alpha_2}{\rho_1\alpha_1}\cos 2\psi_2 & \frac{\rho_2\beta_2}{\rho_1\alpha_1}\sin 2\psi_2 & \frac{\rho_2\alpha_2}{\rho_1\alpha_1}\cos 2\psi_2 F_l & -\frac{\rho_2\beta_2}{\rho_1\alpha_1}\sin 2\psi_2 F_t & 0 & 0 \\ 0 & 0 & -\cos 2\psi_2 F_l & -\frac{\beta_2}{\alpha_2}\sin 2\psi_2 F_t & -\cos 2\psi_2 & \frac{\beta_2}{\alpha_2}\sin 2\psi_2 & \frac{\rho_3\alpha_3}{\rho_2\alpha_2}\cos 2\psi_3 & \frac{\rho_3\beta_3}{\rho_2\alpha_2}\sin 2\psi_3 \\ \sin 2\phi_1 & \frac{\alpha_1}{\beta_1}\cos 2\psi_1 & \frac{\rho_2\alpha_1\beta_2^2}{\rho_1\alpha_2\beta_1^2}\sin 2\phi_2 & -\frac{\rho_2\alpha_1\beta_2}{\rho_1\beta_1^2}\cos 2\psi_2 & -\frac{\rho_2\alpha_1\beta_2^2}{\rho_1\alpha_2\beta_1^2}\sin 2\phi_2 F_l & -\frac{\rho_2\alpha_1\beta_2}{\rho_1\beta_1^2}\cos 2\psi_2 F_t & 0 & 0 \\ 0 & 0 & -\sin 2\phi_2 F_l & \frac{\alpha_2}{\beta_2}\cos 2\psi_2 F_t & \sin 2\phi_2 & \frac{\alpha_2}{\beta_2}\cos 2\psi_2 & \frac{\rho_3\alpha_2\beta_3^2}{\rho_2\alpha_3\beta_2^2}\sin 2\phi_3 & -\frac{\rho_3\alpha_2\beta_3}{\rho_2\beta_2^2}\cos 2\psi_3 \end{bmatrix} \begin{bmatrix} A_1 \\ B_1 \\ A_2 \\ B_2 \\ C_2 \\ D_2 \\ A_3 \\ B_3 \end{bmatrix} = \begin{bmatrix} \cos\phi_0 \\ \sin\phi_0 \\ 0 \\ 0 \\ \cos 2\psi_1 \\ 0 \\ \sin 2\phi_0 \\ 0 \end{bmatrix}$$

**Equations for an incident $S_V$ wave ($B_0 = 1$ and $A_0 = 0$)**

Applying equations 1-8, we obtain:

$\cos\phi_1 A_1 - \sin\psi_1 B_1 + \cos\phi_2 A_2 + \sin\psi_2 B_2 - \cos\phi_2 F_l C_2 + \sin\psi_2 F_t D_2 = -\sin\psi_0$ … (A3.17)

$-\sin\phi_1 A_1 - \cos\psi_1 B_1 + \sin\phi_2 A_2 - \cos\psi_2 B_2 + \sin\phi_2 F_l C_2 + \cos\psi_2 F_t D_2 = \cos\psi_0$ … (A3.18)

$\cos\phi_2 F_l A_2 + \sin\psi_2 F_t B_2 - \cos\phi_2 C_2 + \sin\psi_2 D_2 - \cos\phi_3 A_3 - \sin\psi_3 B_3 = 0$ … (A3.19)

$\sin\phi_2 F_l A_2 - \cos\psi_2 F_t B_2 + \sin\phi_2 C_2 + \cos\psi_2 D_2 - \sin\phi_3 A_3 + \cos\psi_3 B_3 = 0$ … (A3.20)

At z=0

$-\frac{\alpha_1}{\beta_1}\cos 2\psi_1 A_1 + \sin 2\psi_1 B_1 + \frac{\rho_2\alpha_2}{\rho_1\beta_1}\cos 2\psi_2 A_2 + \frac{\rho_2\beta_2}{\rho_1\beta_1}\sin 2\psi_2 B_2$

$+\frac{\rho_2\alpha_2}{\rho_1\beta_1}\cos 2\psi_2 F_l C_2 - \frac{\rho_2\beta_2}{\rho_1\beta_1}\sin 2\psi_2 F_t D_2 = -\sin 2\psi_0$ … (A3.21)

And at z=H

$-\cos 2\psi_2 F_l A_2 - \frac{\beta_2}{\alpha_2}\sin 2\psi_2 F_t B_2 - \cos 2\psi_2 C_2 + \frac{\beta_2}{\alpha_2}\sin 2\psi_2 D_2$

$+\frac{\rho_3\alpha_3}{\rho_2\alpha_2}\cos 2\psi_3 A_3 + \frac{\rho_3\beta_3}{\rho_2\alpha_2}\sin 2\psi_3 B_3 = 0$ … (A3.22)

$\frac{\beta_1}{\alpha_1}\sin 2\phi_1 A_1 + \cos 2\psi_1 B_1 + \frac{\rho_2\beta_2^2}{\rho_1\alpha_2\beta_1}\sin 2\phi_2 A_2 - \frac{\rho_2\beta_2}{\rho_1\beta_1}\cos 2\psi_2 B_2$

$-\frac{\rho_2\beta_2^2}{\rho_1\alpha_2\beta_1}\sin 2\phi_2 F_l C_2 - \frac{\rho_2\beta_2}{\rho_1\beta_1}\cos 2\psi_2 F_t D_2 = \cos 2\psi_0$ … (A3.23)

$-\sin 2\phi_2 F_l A_2 + \frac{\alpha_2}{\beta_2}\cos 2\psi_2 F_t B_2 + \sin 2\phi_2 C_2 + \frac{\alpha_2}{\beta_2}\cos 2\psi_2 D_2$

$\frac{\rho_3\alpha_2\beta_3^2}{\rho_2\alpha_3\beta_2^2}\sin 2\phi_3 A_3 - \frac{\rho_3\alpha_2\beta_3}{\rho_2\beta_2^2}\cos 2\psi_3 B_3 = 0$ … (A3.24)

These equations give rise to matrix form for incident $S_V$-wave

# Karl Zoeppritz

$$\begin{bmatrix} \cos\phi_1 & -\sin\psi_1 & \cos\phi_2 & \sin\psi_2 & -\cos\phi_2 F_l & \sin\psi_2 F_t & 0 & 0 \\ -\sin\phi_1 & -\cos\psi_1 & \sin\phi_2 & -\cos\psi_2 & \sin\phi_2 F_l & \cos\psi_2 F_t & 0 & 0 \\ 0 & 0 & \cos\phi_2 F_l & \sin\psi_2 F_t & -\cos\phi_2 & \sin\psi_2 & -\cos\phi_3 & -\sin\psi_3 \\ 0 & 0 & \sin\phi_2 F_l & -\cos\psi_2 F_t & \sin\phi_2 & \cos\psi_2 & -\sin\phi_3 & \cos\psi_3 \\ -\dfrac{\alpha_1}{\beta_1}\cos 2\psi_1 & \sin 2\psi_1 & \dfrac{\rho_2\alpha_2}{\rho_1\beta_1}\cos 2\psi_2 & \dfrac{\rho_2\beta_2}{\rho_1\beta_1}\sin 2\psi_2 & \dfrac{\rho_2\alpha_2}{\rho_1\beta_1}\cos 2\psi_2 F_l & -\dfrac{\rho_2\beta_2}{\rho_1\beta_1}\sin 2\psi_2 F_t & 0 & 0 \\ 0 & 0 & -\cos 2\psi_2 F_l & -\dfrac{\beta_2}{\alpha_2}\sin 2\psi_2 F_t & -\cos 2\psi_2 & \dfrac{\beta_2}{\alpha_2}\sin 2\psi_2 & \dfrac{\rho_3\alpha_3}{\rho_2\alpha_2}\cos 2\psi_3 & \dfrac{\rho_3\beta_3}{\rho_2\alpha_2}\sin 2\psi_3 \\ \dfrac{\beta_1}{\alpha_1}\sin 2\phi_1 & \cos 2\psi_1 & \dfrac{\rho_2\beta_2^2}{\rho_1\alpha_2\beta_1}\sin 2\phi_2 & -\dfrac{\rho_2\beta_2}{\rho_1\beta_1}\cos 2\psi_2 & -\dfrac{\rho_2\beta_2^2}{\rho_1\alpha_2\beta_1}\sin 2\phi_2 F_l & -\dfrac{\rho_2\beta_2}{\rho_1\beta_1}\cos 2\psi_2 F_t & 0 & 0 \\ 0 & 0 & -\sin 2\phi_2 F_l & \dfrac{\alpha_2}{\beta_2}\cos 2\psi_2 F_t & \sin 2\phi_2 & \dfrac{\alpha_2}{\beta_2}\cos 2\psi_2 & \dfrac{\rho_3\alpha_2\beta_3^2}{\rho_2\alpha_3\beta_2^2}\sin 2\phi_3 & -\dfrac{\rho_3\alpha_2\beta_3}{\rho_2\beta_2^2}\cos 2\psi_3 \end{bmatrix} \begin{bmatrix} A_1 \\ B_1 \\ A_2 \\ B_2 \\ C_2 \\ D_2 \\ A_3 \\ B_3 \end{bmatrix} = \begin{bmatrix} -\sin\psi_0 \\ \cos\psi_0 \\ 0 \\ 0 \\ -\sin 2\psi_0 \\ 0 \\ \cos 2\psi_0 \\ 0 \end{bmatrix}$$